\documentclass[sn-apa, iicol]{sn-jnl}

\usepackage[labelsep=space, labelfont=sf]{caption}
\usepackage[labelfont=bf]{subcaption}

\jyear{2022}%

\theoremstyle{thmstyleone}%
\theoremstyle{thmstyletwo}%

\theoremstyle{thmstylethree}%

\newcommand{\dv}{\mathrm{d}v}                  
\newcommand{\us}{{\mathbf u}_S}

\renewcommand {\div} {\hbox{\rmfamily div}\,}                

\newcommand  {\grad}{\hbox{\rmfamily grad}\,}

\newcommand{\mbfrm}[1]{
	\mbox{%
		\mathversion{bold}%
		$\mathrm{#1}$%
		\mathversion{normal}%
	}%
}

\newcommand{\bfn}{\mbfrm{n}}

\newcommand{\bfw}{\mbfrm{w}}

\newcommand{\diff}{\mathrm{d}}
\newcommand{\alp}{^{\alpha}}
\newcommand{\ualp}{_{\alpha}}
\newcommand{\bet}{^{\beta}}
\newcommand{\ubet}{_{\beta}}

\newcommand{\bx}{\mathbf{x}}

\newcommand{\bs}{\boldsymbol}
\newcommand{\mbf}{\mathbf}
\newcommand{\mrm}{\mathrm}
\newcommand{\vs}{\overset{\prime}{\bx}_S}

\newcommand{\vbeta}{\overset{\prime}{\bx}_{\beta}}

\newcommand{\dgamma}{\dot{\gamma}}
\newcommand{\rh}{_{rh}}
\newcommand{\eff}{_{eff}}
\newcommand{\muinf}{\mu_{\infty}}
\newcommand{\muz}{\mu_{0}}

\begin{document}

\title[ ]{A continuum mechanical porous media model for vertebroplasty: Numerical simulations and experimental validation}


\author*[1]{\fnm{Zubin} \sur{Trivedi}}\email{zubin.trivedi@imsb.uni-stuttgart.de}

\author[2]{\fnm{Dominic} \sur{Gehweiler}}\email{dominic.gehweiler@aofoundation.org}

\author[2]{\fnm{Jacek K.} \sur{Wychowaniec}}\email{jacek.wychowaniec@aofoundation.org}

\author[3]{\fnm{Tim} \sur{Ricken}}\email{tim.ricken@isd.uni-stuttgart.de}

\author[2]{\fnm{Boyko} \sur{Gueorguiev}}\email{boyko.gueorguiev@aofoundation.org}

\author[4,5]{\fnm{Arndt} \sur{Wagner}}\email{arndt.wagner@mechbau.uni-stuttgart.de}

\author[1,5]{\fnm{Oliver} \sur{R\"ohrle}}\email{roehrle@simtech.uni-stuttgart.de}

\affil[1]{\orgdiv{Institute for Modelling and Simulation of Biomechanical Systems}, \orgname{University of Stuttgart}, \orgaddress{\street{Pfaffenwaldring 5a}, \postcode{70569} \city{Stuttgart}, \country{Germany}}}

\affil[2]{\orgname{AO Research Institute (ARI)}, \orgaddress{\street{Clavadelerstrasse 8},  \postcode{7270} \city{Davos}, \country{Switzerland}}}

\affil[3]{\orgdiv{Institute of Structural Mechanics and Dynamics in Aerospace Engineering}, \orgname{University of Stuttgart}, \orgaddress{\street{Pfaffenwaldring 27}, \postcode{70569} \city{Stuttgart}, \country{Germany}}}

\affil[4]{\orgdiv{Institute of Applied Mechanics (CE)}, \orgname{University of Stuttgart}, \orgaddress{\street{Pfaffenwaldring 7}, \postcode{70569} \city{Stuttgart}, \country{Germany}}}

\affil[5]{\orgdiv{Stuttgart Center for Simulation Science (SC SimTech)}, \orgname{University of Stuttgart}, \orgaddress{\street{Pfaffenwaldring 5a}, \postcode{70569} \city{Stuttgart}, \country{Germany}}}


\abstract{The outcome of vertebroplasty is hard to predict due to its dependence on complex factors like bone cement and marrow rheologies. Cement leakage could occur if the procedure is done incorrectly, potentially causing adverse complications. A reliable simulation could predict the patient-specific outcome preoperatively and avoid the risk of cement leakage. Therefore, the aim of this work was to introduce a computationally feasible and experimentally validated model for simulating vertebroplasty. The developed model is a multiphase continuum-mechanical macro-scale model based on the Theory of Porous Media. The related governing equations were discretized using a combined Finite Element - Finite Volume approach by the so-called Box discretization. Three different rheological upscaling methods were used to compare and determine the most suitable approach for this application. For validation, a benchmark experiment was set up and simulated using the model. The influence of bone marrow and parameters like permeability, porosity, etc., was investigated to study the effect of varying conditions on vertebroplasty. The presented model could realistically simulate the injection of bone cement in porous materials when used with the correct rheological upscaling models, of which the semi-analytical averaging of the viscosity gave the best results. The marrow viscosity is identified as the crucial reference to categorize bone cements as ‘high-' or ‘low-' viscosity in the context of vertebroplasty. It is confirmed that a cement with higher viscosity than the marrow ensures stable development of the injection and a proper cement interdigitation inside the vertebra.}

\keywords{Vertebroplasty, Bone cement, Porous media, Non-Newtonian}



\maketitle

\section*{Statements and Declarations}

\noindent \textbf{Funding/Support Statement}\\
Funded by the Deutsche Forschungsgemeinschaft (DFG, German Research Foundation) – Project Number 327154368 – SFB 1313. We also acknowledge the support by the Stuttgart Center for Simulation Science (SimTech). J.K.W.~acknowledges the European Union’s Horizon 2020 (H2020-MSCA-IF-2019) research and innovation programme under the Marie Skłodowska-Curie grant agreement 893099 — ImmunoBioInks.\\

\noindent \textbf{Acknowledgements} \\
We heartily thank Johannes Hommel, Martin Schneider, and Prof.~Rainer Helmig from the Department of Hydromechanics and Modelling of Hydrosystems at the University of Stuttgart for providing help with pore-network modelling and implementing the Box discretization. Thanks are also extended to Christian Bleiler and Yesid Narvaez from the Institute for Modelling and Simulation of Biomechanical Systems at the University of Stuttgart for proofreading the manuscript.\\

\noindent \textbf{Author Contributions} \\
Conceptualization: 	Arndt Wagner, Tim Ricken, Boyko Gueorguiev-R\"uegg, Oliver R\"ohrle; Methodology: Zubin Trivedi, Dominic Gehweiler, Jacek K. Wychowaniec; Formal analysis and investigation: Zubin Trivedi, Dominic Gehweiler, Jacek K. Wychowaniec, Arndt Wagner; Writing - original draft preparation: Zubin Trivedi; Writing - review and editing: Dominic Gehweiler, Jacek K. Wychowaniec, Arndt Wagner; Funding acquisition: Arndt Wagner, Tim Ricken, Oliver R\"ohrle; Resources: Boyko Gueorguiev-R\"uegg, Oliver R\"ohrle; Supervision: Arndt Wagner, Tim Ricken, Boyko Gueorguiev-R\"uegg, Oliver R\"ohrle. All authors have read and agreed to the published version of the manuscript. \\

\noindent \textbf{Conflict of Interest}\\
The authors have no conflicts of interest to disclose in relation to this article.\\

\noindent \textbf{Data Availability} \\
All data needed to evaluate the conclusions and used in the graphs are available at: https://doi.org/10.18419/darus-3146.

\clearpage

\section{Introduction}
\label{sec:intro}
Annually, more than 1.4 million clinical vertebral fractures occur worldwide \citep{johnell_estimate_2006}. Vertebral fractures lead to back pain, loss of height, immobility, reduced pulmonary function, and increased mortality. Vertebroplasty is a common medical procedure for treating and preventing vertebral fractures. In this procedure, a so-called ``bone cement" is injected into the porous interior of the vertebra, where it undergoes curing, providing additional structural strength to the vertebra \citep{jensen_percutaneous_1997}. Vertebroplasty provides quick pain relief in 80-90 \% of the cases \citep{mcgraw_prospective_2002}. Generally, the practitioners use haptic feedback and X-ray imaging to guide the cement injection. However, this is difficult due to several factors, including the non-constant viscosity of the bone cement owing to its non-Newtonian nature and the curing process. The bone marrow in the vertebra is also a non-Newtonian fluid. Severe complications could occur if the cement leaks into the blood vessels or the spinal canal due to improper injection, e.g.~pulmonary embolism or paralysis \citep{bernhard_asymptomatic_2003, ratliff_root_2001}. In this regard, a reliable simulation of the procedure could help practitioners predict the risks and determine the safe ranges for operating parameters like injection pressure, cement viscosity, curing time, flow rate, etc., specific to each case. The decrease in reliance on X-ray imaging would also reduce the patient's radiation exposure. 

In the past, various approaches have been used for simulating vertebroplasty. Theoretical approaches, e.g. \citep{bohner_theoretical_2003}, usually have limited use-case because many idealizing assumptions are necessary. The earliest works using computational approaches were limited to small two-dimensional segments and assumed the porous structure to be a bundle of capillary tubes, e.g.~in \citep{beaudoin_finite_1991}. With improvement in computational capabilities, more complicated models appeared, such as branching pipe networks \citep{lian_biomechanical_2008}, Computational Fluid Dynamics (CFD) models \citep{teo_preliminary_2007,landgraf_modelling_2015} or Lattice-Boltzmann models \citep{zeiser_pore-scale_2008}. The main drawback in these models is that a pore-scale flow resolution inside a complicated porous structure quickly becomes computationally expensive. Therefore, most pore-scale models simulate flow only over a small part or a representative segment of the vertebra. For a simulation regarding the entire vertebra, a macro-scale model is a more feasible option. In general, there is substantial research on the mathematical modelling of such multiphase porous media flow problems at the macro-scale; however, most of it is in the context of groundwater flow. It is possible to extend and adapt such models to vertebroplasty, but a major challenge is to upscale the viscosities of the non-Newtonian fluids for use in the macro-scale flow equations. \cite{widmer_soyka_numerical_2013} developed such a macro-scale model using a continuum length scale Reynolds number for upscaling the viscosity. However, they employed a front-tracking Volume of Fluid method that assumes a sharp cement-marrow interface \citep{widmer_mixed_2011}, which may not always be the case. In this regard, the Theory of Porous Media (TPM) \citep{ehlers_foundations_2002, ehlers_challenges_2009} provides a solid framework for implementing a multiphase continuum-mechanical macro-scale model to simulate the fluid flow in a porous medium as well as the deformations in the porous medium, without needing to idealize the interface between the fluids. The application of TPM application extends to, for example, the biomechanics of soft tissues such as liver ~\citep{Ricken.2019}, brain ~\citep{Ehlers.2022}, and cartilage~\citep{Wang.2018}. TPM has previously also been used to develop a preliminary model for vertebroplasty \citep{bleiler_multiphasic_2015}. 

The aim of this work was to develop a computationally feasible model based on TPM, and to extend it beyond previous studies \citep{bleiler_multiphasic_2015} for better suitability of simulating vertebroplasty by including a more appropriate choice of model variables and constitutive equations for non-Newtonian rheology, investigating suitable rheology upscaling models, and numerically treating the governing equations by a combined Finite Element - Finite Volume approach using the Box discretization to address the different natures of the fluid flow and the solid deformation problem in a stable framework. Furthermore, the work aims to validate the model using a benchmark problem, which is investigated experimentally as well as simulated computationally; and carry out further simulations to determine the influence of various parameters, with specific regard to avoiding cement leakage.  

\section{Materials and Methods}
\label{sec:methods}

\subsection{Modelling approach}
\label{sec:numerical_model}

\subsubsection{Multiphase modelling based on the Theory of Porous Media}
\label{sec:TPM}

\begin{figure}[htbp]
	\begin{subfigure}[t]{0.95\linewidth}
		\centering
		\includegraphics[width=\textwidth]{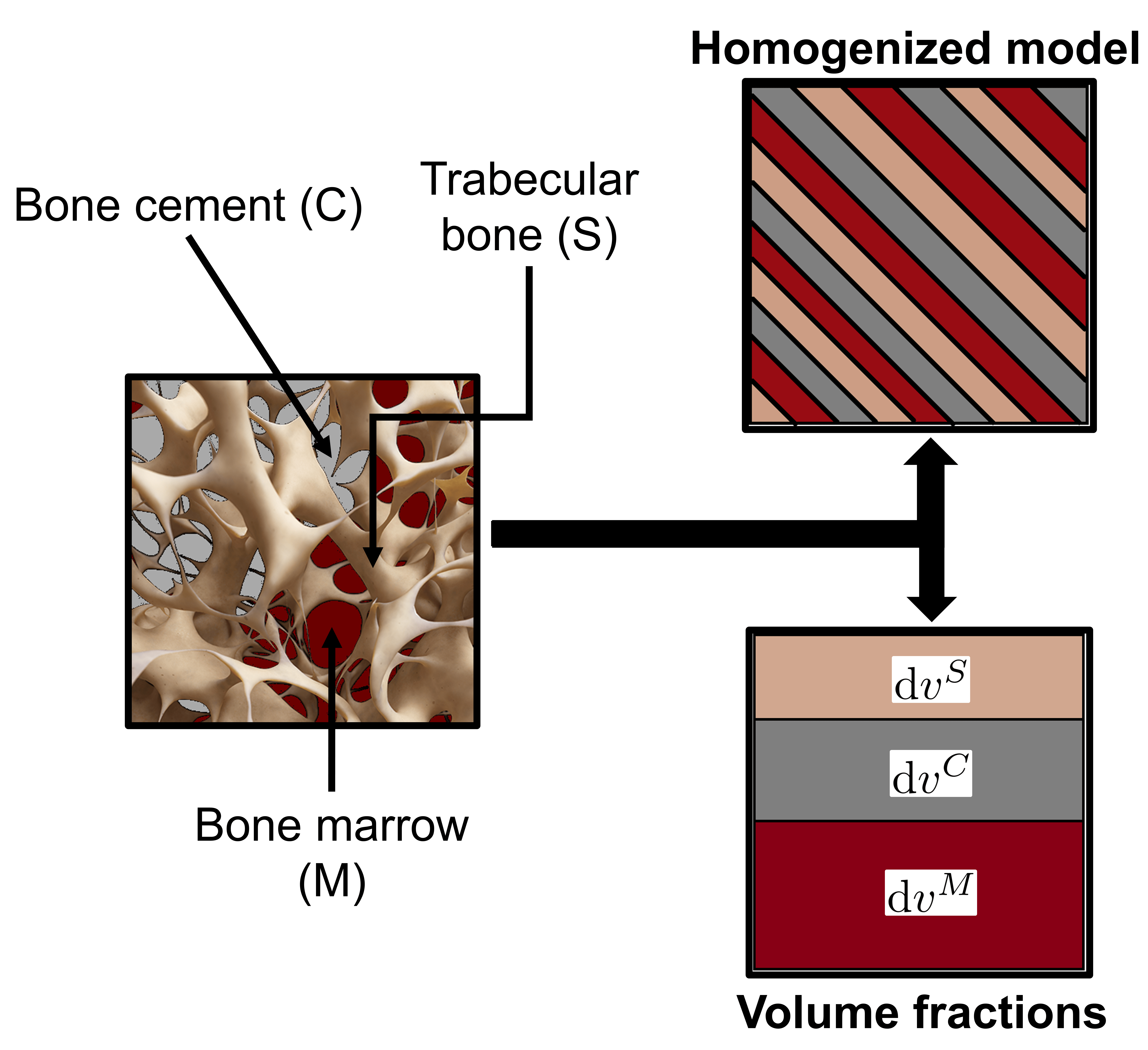}
		\caption{}
		\label{fig:homogenization}
	\end{subfigure}
	\begin{subfigure}[b]{\linewidth}
		\centering
		\includegraphics[width=\textwidth]{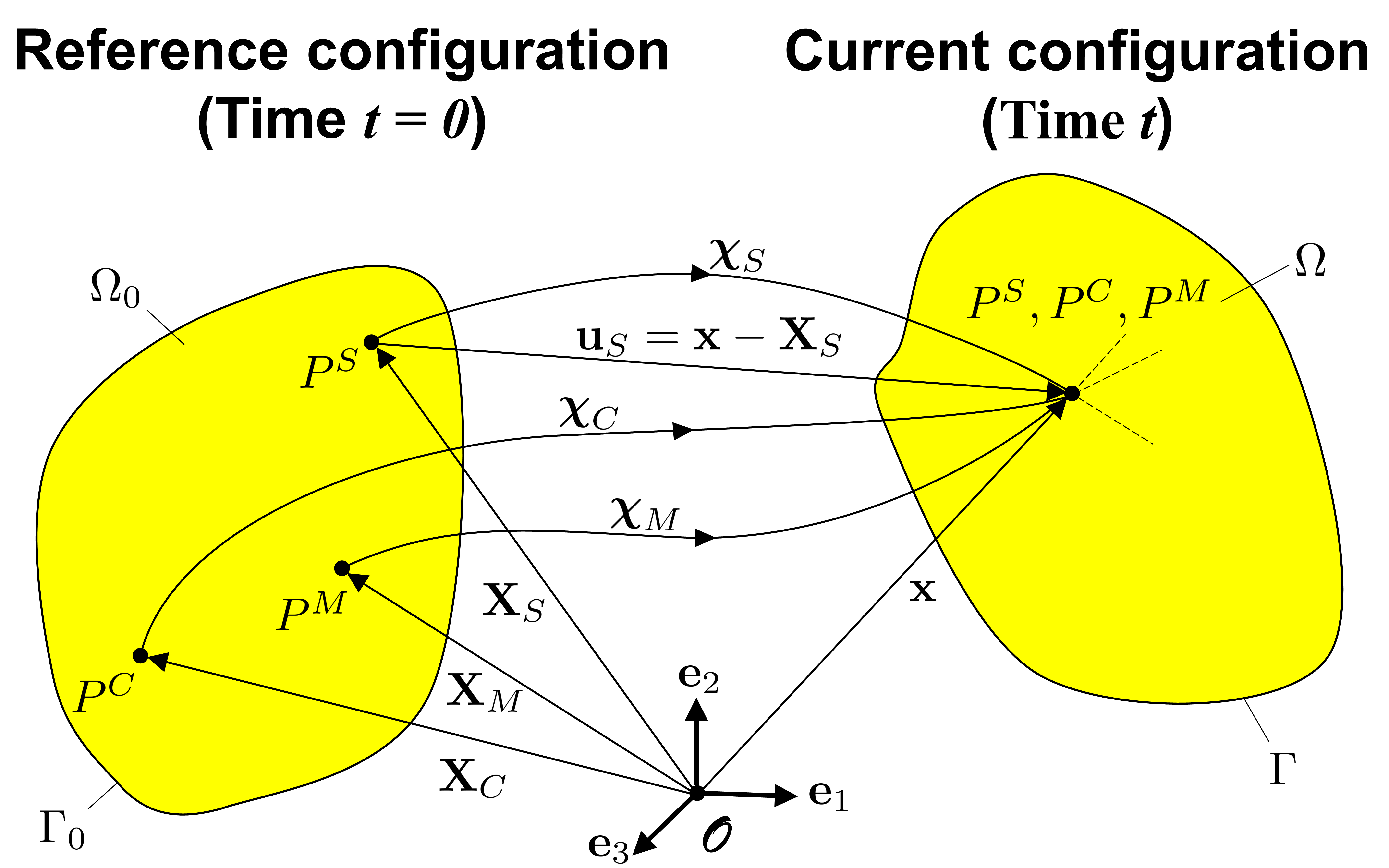}
		\caption{}
		\label{fig:continuum}
	\end{subfigure}
	\caption{\textbf{(a)} Homogenization over a Representative Elementary Volume (REV) of the vertebra interior \textbf{(b)} Kinematics of a multiphase continuum body with three phases: solid bone (S), bone cement (C), and bone marrow (M)}
\end{figure}

The Theory of Porous Media (TPM) provides a framework for modelling the mechanics of a multiphase continuum at the macro-scale \citep{ehlers_foundations_2002, ehlers_challenges_2009}. It originates from the Theory of Mixtures \citep{bowen_porous_1984}, where the pore-scale properties of the phases are homogenized over a so-called Representative Elementary Volume (REV), such that the macroscopic properties of this volume are representative of the entire domain. In our case, the overall aggregate contains three phases: the solid trabecular bone $\varphi^S$, the bone cement $\varphi^C$, and the bone marrow $\varphi^M$, such that $\varphi = \bigcup_{\alpha} \varphi\alp = \varphi^S \, \cup \, \varphi^F = \varphi^S \, \cup \, \varphi^C \,\cup \, \varphi^M $. Here, $\alpha = \{S, C, M\}$ and $\varphi^F = \bigcup_{\beta} \varphi^\beta$, where the fluid phases $\beta = \{C, M\}$ are immiscible since the bone cement is hydrophobic. In the TPM, this is further extended by including the individual amount of each phase using the concept of volume fractions. This is shown in Figure \ref{fig:homogenization}. The volume fraction $n \alp = \dv\alp/\dv$, where $\dv\alp$ is the volume of the phase $\dv\alp$ and $\dv$ is the total volume of the REV, sum up to one, i.e. $\sum \ualp n \alp = 1$. Thus, the medium does not have any void spaces. Similarly, the saturations $s^{\beta} = n^{\beta}/{n^F}$ of the fluid phases are introduced. 

Under loading conditions, such a system undergoes a change in its state with time. The state at the current time $t$ is called the current configuration, whereas that at the reference time $t = 0$ is the reference configuration. Each spatial point with a position vector $\bx$ in the current configuration is occupied by all three phases (superimposed continua). Each of these three phases may originate from a different point with position vector $\mbf{X} \ualp$ in the reference configuration, depending on their respective placement function $\boldsymbol{\chi} \ualp$, such that $\bx = \boldsymbol{\chi} \ualp (\mbf{X} \ualp, t)$. The material time derivative of each phase is defined as $\overset{\prime}{\bx} \ualp = \mrm{d}(\boldsymbol{\chi} \ualp (\mbf{X} \ualp, t))/\mrm{d} t$. The deformation of the bone is then defined in a Lagrangian setting as $\mbf{u} \ualp = \bx - \mbf{X} \ualp$, while the movement of the fluid phases is described in a `modified' Eulerian setting by their seepage velocities, given as $\mbf{w}_{\beta} = \vbeta - \vs$. The kinematics are shown in Figure \ref{fig:continuum}.

These variables are governed by a set of balance equations, viz., the mass balance and the momentum balance equations for each of the phases. These are obtained after applying the assumptions that all phases are materially incompressible, no mass exchange occurs between the phases, the porous medium is fully saturated, and the constituents are non-polar (stress tensors are symmetric). Additionally, quasi-static processes are assumed, body forces are neglected, and all processes are considered isothermal. This yields the volume balance law for the phases as 
\begin{equation}
	(n \alp)' \ualp + n \alp \div \overset{\prime}{\bx}\ualp = 0\,.
\end{equation}
For the solid, this becomes simply
\begin{equation}
	(n^S)'_S + n^S \div \vs = 0 \quad \longrightarrow  \quad n^S = n^S_{0S}\, (\det \mbf{F}_S)^{-1}\,,
	\label{eq:vol_balance_solid}
\end{equation}
where $n^S_{0S}$ is the initial volume fraction and $\mbf{F}_S = {\mrm{d} \bx }/{\mrm{d} \mbf{X}_S}$ is the deformation gradient. For fluids, the volume balance law in the modified Eulerian setting becomes
\begin{equation}
	(n^\beta)'_S + \div(n^\beta \mbf{w}_\beta) + n^\beta \div \vs = 0\,.
	\label{eq:vol_balance_fluids}
\end{equation}
Additionally, we have the overall momentum balance given as 
\begin{equation}
	\div \boldsymbol{\sigma} = \mbf{0}\,,
	\label{eq:mtm_balance_aggr}
\end{equation}
where $\boldsymbol{\sigma} = \boldsymbol{\sigma}^S_E - p \mbf{I}$\,. The solid extra stress $\boldsymbol{\sigma}^S_E$ is related to the deformation by constitutive relations, and $p$ is the hydrostatic pore pressure. For the considered application, the capillary number is large ($Ca \gg 10^{-3}$), implying that the capillary forces are negligible compared to the viscous forces. Therefore, we neglect the pressure difference between the two fluids arising due to capillary effects and assume the same pore pressure $p$ for both fluid phases. The resulting stress of the solid skeleton is obtained by $\boldsymbol{\sigma}^S = \boldsymbol{\sigma}^S_E - n^S p \mbf{I}$, and the fluid stress is given by $\boldsymbol{\sigma}^\beta = - n^\beta p \mbf{I}$. Here, we neglect the fluid extra stresses $\boldsymbol{\sigma}^\beta_E$ arising from viscous effects compared to the local interaction forces, as can be concluded from dimensional analysis \citep{markert_constitutive_2007}.

Furthermore, the Darcy flow equations are used to relate the seepage velocities of the fluid phases to the pressure gradient. These equations can be derived in the framework of the TPM by applying constitutive relations to the individual momentum balances of the fluid phases $\varphi^\beta$. These are then written as 
\begin{align}
	n^C \mbf{w}_C &= - \frac{\kappa^C_r \mbf{K}^S}{\mu^{CR}} \grad p \,,
	\label{eq:darcyc}\\
	n^M \mbf{w}_M &= - \frac{\kappa^M_r \mbf{K}^S}{\mu^{MR}} \grad p \,,
	\label{eq:darcym}
\end{align}
where $\mbf{K}^S$ is the intrinsic permeability of the porous medium, $\kappa^\beta_r$ is the relative permeability of the fluid phases arising due to the resistance from the other fluid, and $\mu^{\beta R}$ is the dynamic viscosity. For a detailed overview of the TPM, the reader is referred to the works of Ehlers \citep{ehlers_foundations_2002, ehlers_challenges_2009}.

\subsubsection{Numerical treatment}
\label{sec:implementation}
The governing equations are solved over the entire domain $\Omega$ of the porous medium using the primary variables solid displacement $\us$, pore pressure $p$, and bone marrow saturation $s^M$. Since we neglected the capillary forces for our intended application, we used saturation as one of the primary variables, making the formulation independent of the capillary pressure - saturation relation, in contrast to one with two pore pressures as primary variables (like in \cite{bleiler_multiphasic_2015}). The volume balance for the solid (Equation \ref{eq:vol_balance_solid}) is implemented directly such that it is satisfied at all points, while those for the fluids (Equation \ref{eq:vol_balance_fluids}) and the momentum balance (Equation \ref{eq:mtm_balance_aggr}) are satisfied weakly, i.e., only their weighted integrals over the domain must be zero. This is written as
\begin{align}
	&\int_\Omega [\div \boldsymbol{\sigma}^S_E - \grad p] \cdot \delta \us \,\diff \Omega = 0 \label{eq:weak_mtm} \\
	&\int_\Omega [(n^C)'_S + \div (n^C \bfw_C) + n^C \div \vs] \, \delta p \, \diff \Omega = 0 \label{eq:weak_cement}\\
	&\int_\Omega [(n^M)'_S + \div (n^M \bfw_M) + n^M \div \vs] \, \delta s^M \, \diff \Omega \nonumber \\ &= 0 \label{eq:weak_marrow}
\end{align}
where $\delta\us$, $\delta p$, and $\delta s^M$ are test functions. Applying the Gaussian divergence theorem, the surface integrals for the Neumann boundary conditions appear via 
\begin{align}
	&\int_\Gamma (\boldsymbol{\sigma}_E^S \, \delta \mbf{u}_S) \cdot \hat{\mbf{n}} \,  \diff \Gamma - \int_\Omega \boldsymbol{\sigma}_E^S \cdot \grad \delta \mbf{u}_S \, \diff \Omega \nonumber \\ 
	&- \int_\Omega \grad p \cdot \delta \mbf{u}_S \, \diff \Omega = 0 \label{eq:weak_mtm_2} \\ 
	&\int_\Gamma n^C \bfw_C \cdot \hat{\bfn} \, \delta p \, \diff \Gamma + \int_\Omega (n^C)'_S \, \delta p \, \diff \Omega \nonumber \\ 
	& - \int_\Omega n^C \bfw_C\, \grad \delta p \, \diff \Omega + \int_\Omega n^C  \div \vs \, \delta p \, \diff \Omega = 0   \label{eq:weak_cement_2} \\ 
	&\int_\Gamma n^M \bfw_M \cdot \hat{\bfn} \, \delta s^M \, \diff \Gamma + \int_\Omega (n^C)'_S \, \delta s^M \, \diff \Omega \nonumber \\ 
	&- \int_\Omega n^M \bfw_M\, \grad \delta s^M \, \diff \Omega  + \int_\Omega n^M  \div \vs \, \delta s^M \, \diff \Omega \nonumber \\ &= 0   \label{eq:weak_marrow_2}
\end{align}
where $\hat{\bfn}$ is the surface normal vector. Equation \ref{eq:weak_mtm_2} is spatially discretized and solved using the Bubnov-Galerkin method by the Finite Element discretization. In Equations \ref{eq:weak_cement_2} and \ref{eq:weak_marrow_2}, the volume integral terms consist of a temporal evolution term, a flow term, and a contribution from the solid deformation, therefore requiring both temporal and spatial discretization. For the temporal discretization, we used the Implicit Euler method. For flow problems, especially those hyperbolic by nature, it is known that discretization using the Bubnov-Galerkin Finite Element method leads to numerical instabilities \citep{zienkiewicz_finite_1978}. Therefore, we used the Box spatial discretization scheme and applied mass-lumping for the temporal evolution term. This yields a setting which behaves similar to the Finite Volume discretization. The flow equations were then solved using the fully-upwind Galerkin scheme. More details about the Box discretization can be found in \cite{huber_node-centered_2000}. For the last remaining term, i.e., the contribution from the solid deformation, we again used the Finite Element discretization. The system was discretized using eight-noded hexahedral elements with linear shape functions for all primary variables. The numerical implementation of the model was done on the monolithic solver framework PANDAS \footnote{\textbf{P}orous media \textbf{A}daptive \textbf{N}onlinear finite-element solver based on \textbf{D}ifferential \textbf{A}lgebraic \textbf{S}ystems (http://www.get-pandas.com)}

\subsubsection{Constitutive models}
\label{sec:constitutive}
The governing balance relations need to be supplemented by constitutive relations to include the specific material behaviour of the constituents. For the solid part, i.e., the trabecular bone, the relationship between the stress and the strain is described using the linear Hookean law 
\begin{equation}
	\bs{\sigma}^S_E = 2 \mu_{Lam\acute{e}}\, \bs{\epsilon}_S + \lambda_{Lam\acute{e}} (\bs{\epsilon}_S : \mbf{I}) \mbf{I}
\end{equation}
where $\bs{\sigma}^S_E$ is the solid extra stress, $\bs{\epsilon}_S$ is the strain tensor, $\mu_{Lam\acute{e}}$ and $\lambda_{Lam\acute{e}}$ are the Lam\'{e} material parameters. 

For the two fluids, constitutive relations are required for the change in viscosity with shear rate. We used the Carreau model for this purpose, given for a fluid $\beta$ as 
\begin{equation}
	\mu^{\beta R} = \muinf\bet + (\muz\bet - \muinf \bet) \, [1 + (\lambda^{\beta}\rh \, \dgamma\bet)^{2}]^{\frac{n\bet\rh - 1}{2}}
	\label{eq:carreau}
\end{equation}
where $\muz\bet$ is the viscosity limit at zero shear rate, $\muinf\bet$ is the viscosity limit at very high shear rates, $\lambda\bet\rh$ is the reciprocal of the shear rate where the behaviour transitions to non-Newtonian, $n\bet\rh$ is the flow behaviour index, and $\dgamma\bet$ is the shear rate. Note that the superscript `$\beta$' stands for the fluid phase ($C$ or $M$ for cement or marrow respectively) and the subscript `$rh$' stands for `rheological parameter' to avoid confusion with similar symbols used in other places. 

The viscosity discussed up to this point is a pore-scale quantity. Unlike for a Newtonian fluid, the pore-scale viscosity cannot be directly used in the macro-scale equations (Equations \ref{eq:darcyc}, \ref{eq:darcym}) for a non-Newtonian fluid because the shear rates, and hence the viscosities, vary with the geometries of the pores through which the fluid flows. The viscosity at the macro-scale must be a representative average of the pore-scale viscosities obtained through upscaling. Here, we compared three different upscaling models to evaluate their applicability to model vertebroplasty at the macro-scale. The first two, namely the Cannella model \citep{cannella_prediction_1988} and the Hirasaki and Pope model \citep{hirasaki_analysis_1974}, are based on the capillary-tube-bundle model and semi-empirical by nature. In both approaches, the effective shear rate is obtained from the equation given as
\begin{equation}
	\dgamma\bet\eff = \mathbb{C}\,\bigg[\frac{3n\bet\rh+1}{4n\bet\rh}\bigg]^{\frac{n\bet\rh}{n\bet\rh-1}}\,\bigg[4 \lvert\bfw\ubet\rvert \sqrt{\frac{n\bet}{8 \kappa_r\bet K^S}} \bigg]
	\label{eq:rheo_upscaling}
\end{equation}
where $\dgamma\bet\eff$ is the effective upscaled shear rate,  and $\mathbb{C}$ is a semi-empirical constant, the value of which varies depending on the internal geometry of the porous medium. The Cannella model uses the value $\mathbb{C}= 6.0$, while the Hirasaki and Pope model uses $\mathbb{C}= 0.69$. The effective shear rate is then plugged in Equation \ref{eq:carreau} to obtain the effective viscosity at the macro-scale. Alternatively, the third upscaling method under consideration was the average viscosity model by \cite{eberhard_determination_2019}, which is semi-analytical by nature in contrast to the previous two. This approach analytically computes the average viscosity of the fluid flowing through a representative pore channel of characteristic radius $R_{char}$. This characteristic radius $R_{{char}}$ is determined from the pore size distribution of the porous medium. The details about the derivations of the equations and their analytical solution can be found in \cite{eberhard_determination_2019}. 

Finally, constitutive relations are required for the relative permeabilities. Relative permeability is a factor applied to account for the reduction in permeability due to mutual hindrance caused by the fluid phases. Hence, the constitutive relation is usually a function of saturation. In this work, the Brooks-Corey model \citep{brooks_corey_1964} was used, given as 
\begin{align}
	\begin{split}
		\kappa^C_r &= (1 - s^M)^2 \bigg[1 - (s^M)^{\frac{2+\lambda_{bc}}{\lambda_{bc}}} \bigg] \\
		\kappa^M_r &= (s^M)^{\frac{2+3\lambda_{bc}}{\lambda_{bc}}}
	\end{split}
	\label{eq:brooks-corey}
\end{align}
where $\lambda_{bc}$ is the uniformity parameter. A large value means that pore sizes in the porous medium do not strongly vary, while a small value represents non-uniformity in the pore sizes. 

\subsection{Validation with benchmark problem}
\label{sec:methods_validation}

\begin{figure*}[htbp]
	\begin{subfigure}[t]{0.5\textwidth}
		\centering
		\includegraphics[width=\textwidth]{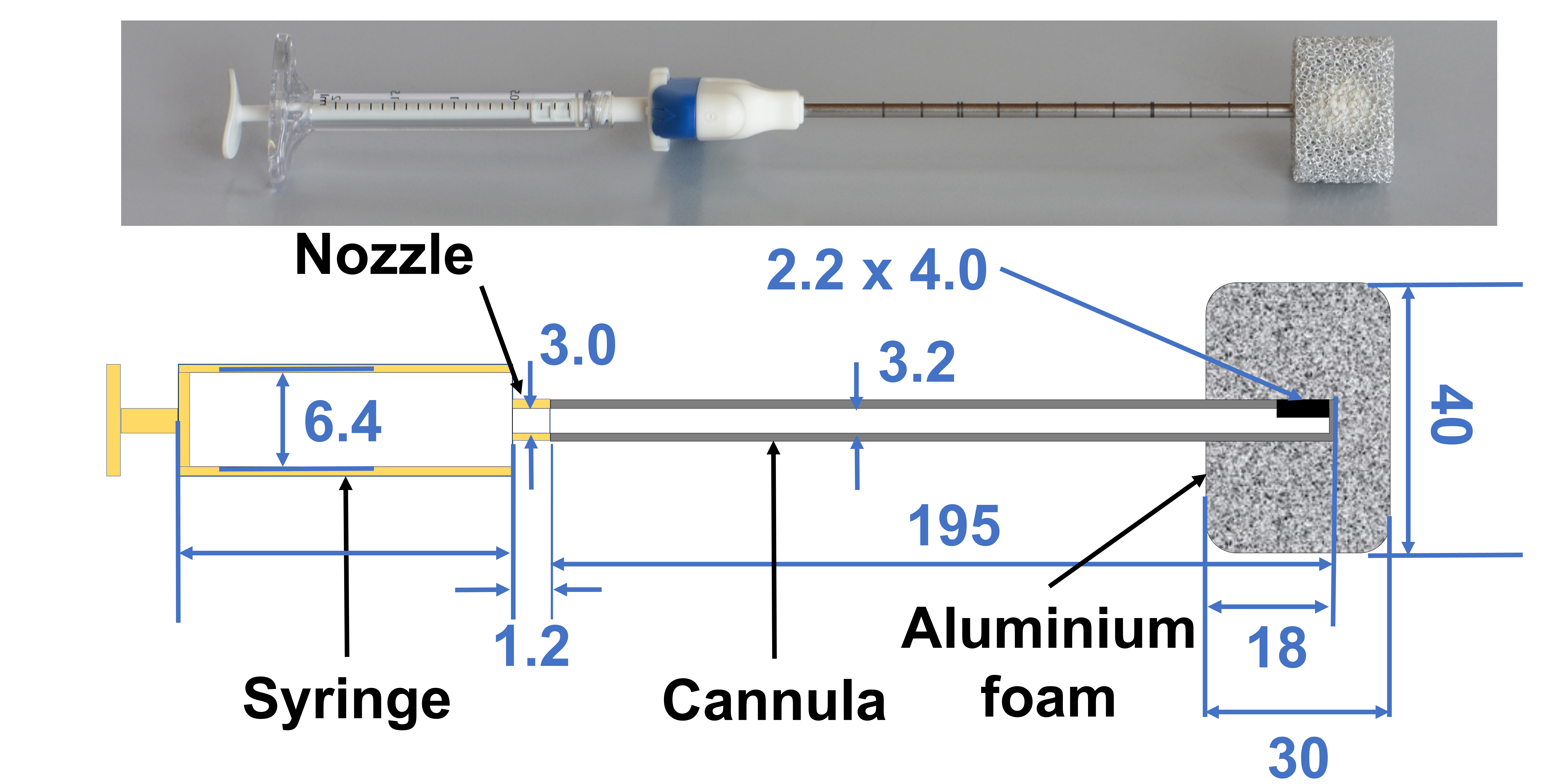}
		\caption{}
		\label{fig:schematic}
	\end{subfigure}
	\hfill
	\begin{subfigure}[t]{0.5\textwidth}
		\centering
		\includegraphics[width=\textwidth]{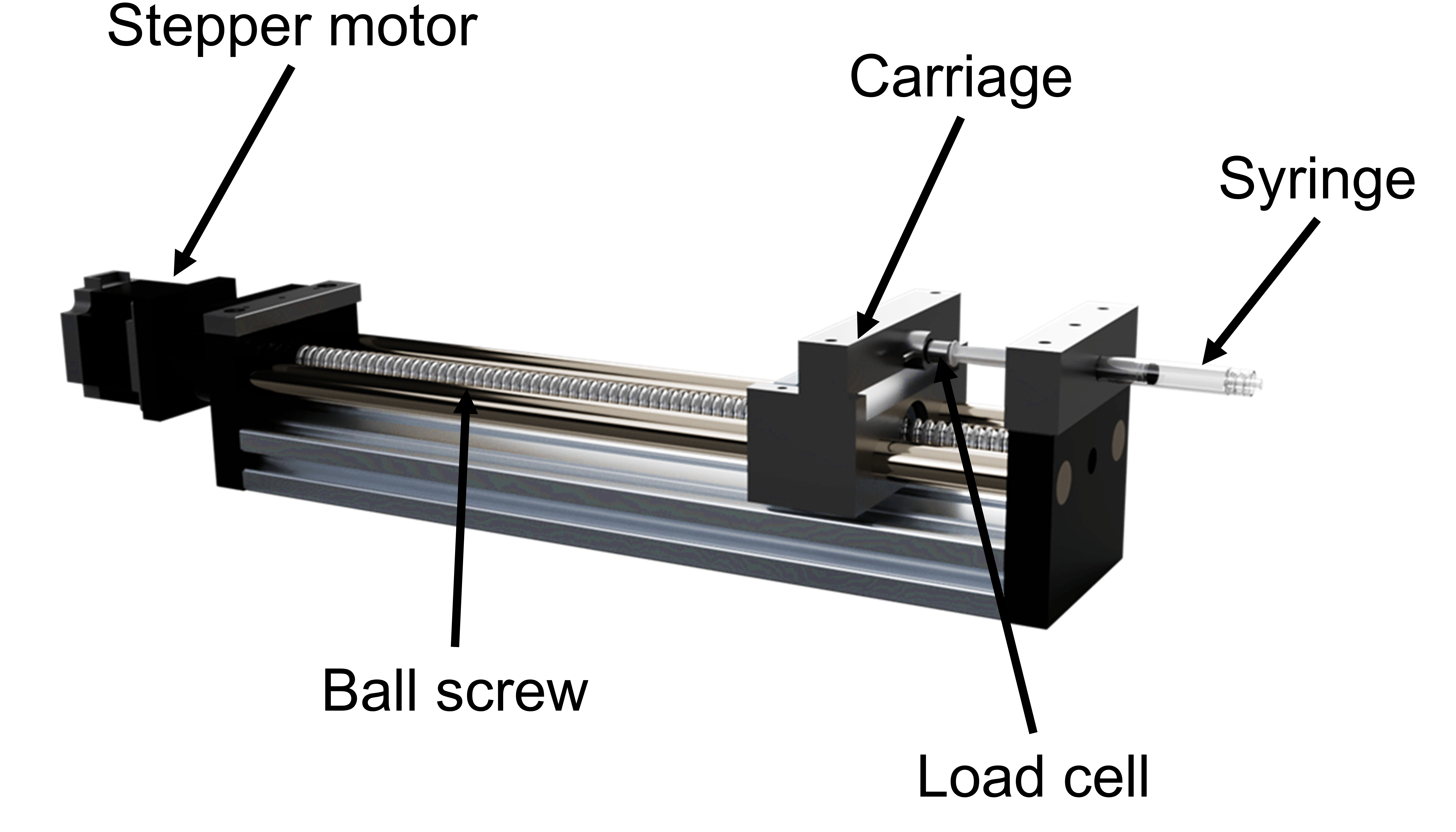}
		\caption{}
		\label{fig:exp_setup_injector}
	\end{subfigure}
	\begin{subfigure}[b]{0.5\textwidth}
		\centering
		\includegraphics[width=\textwidth]{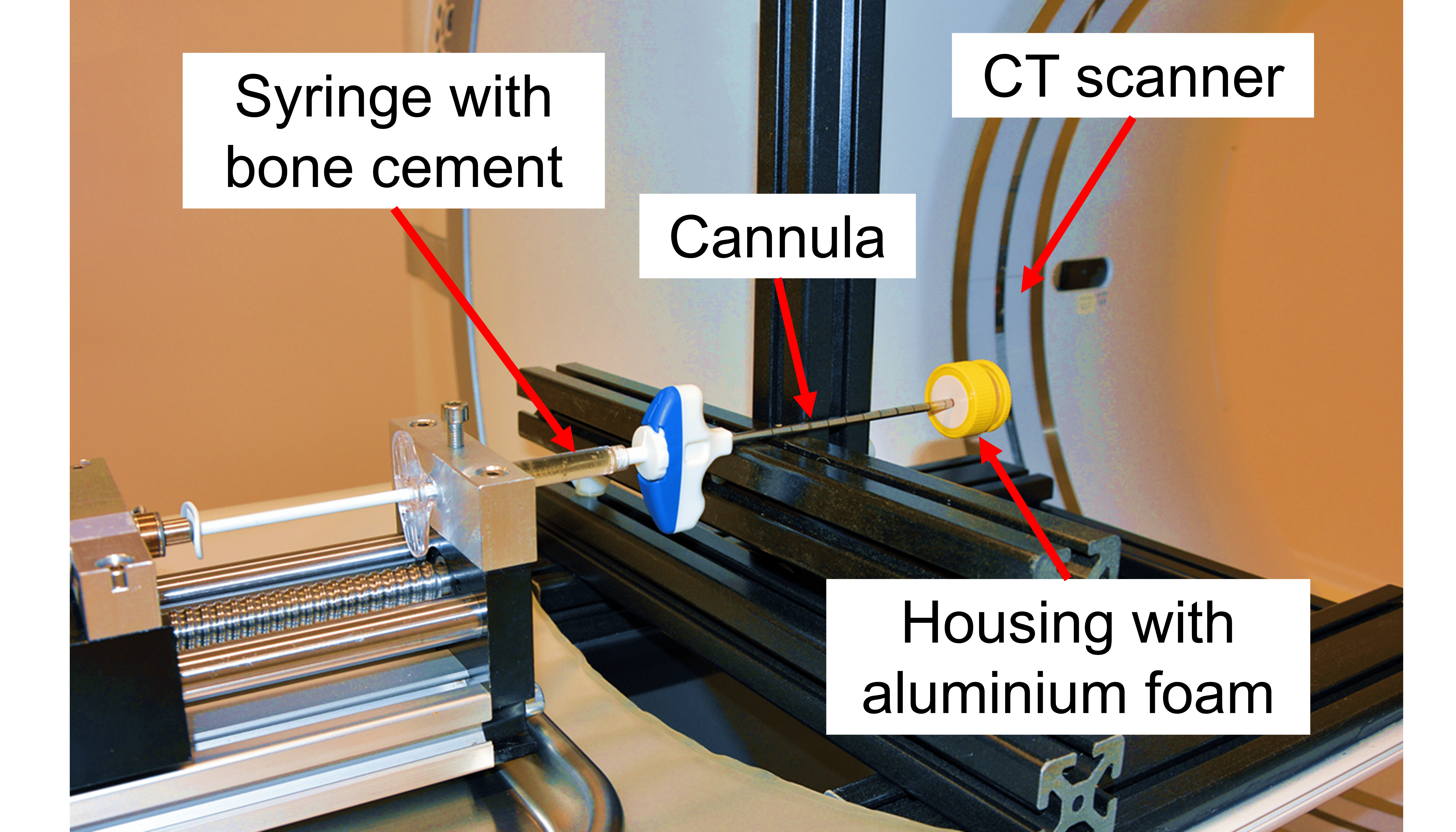}
		\caption{}
		\label{fig:exp_setup_ct}
	\end{subfigure}
	\hfill
	\begin{subfigure}[b]{0.5\textwidth}
		\centering
		\includegraphics[width=\textwidth]{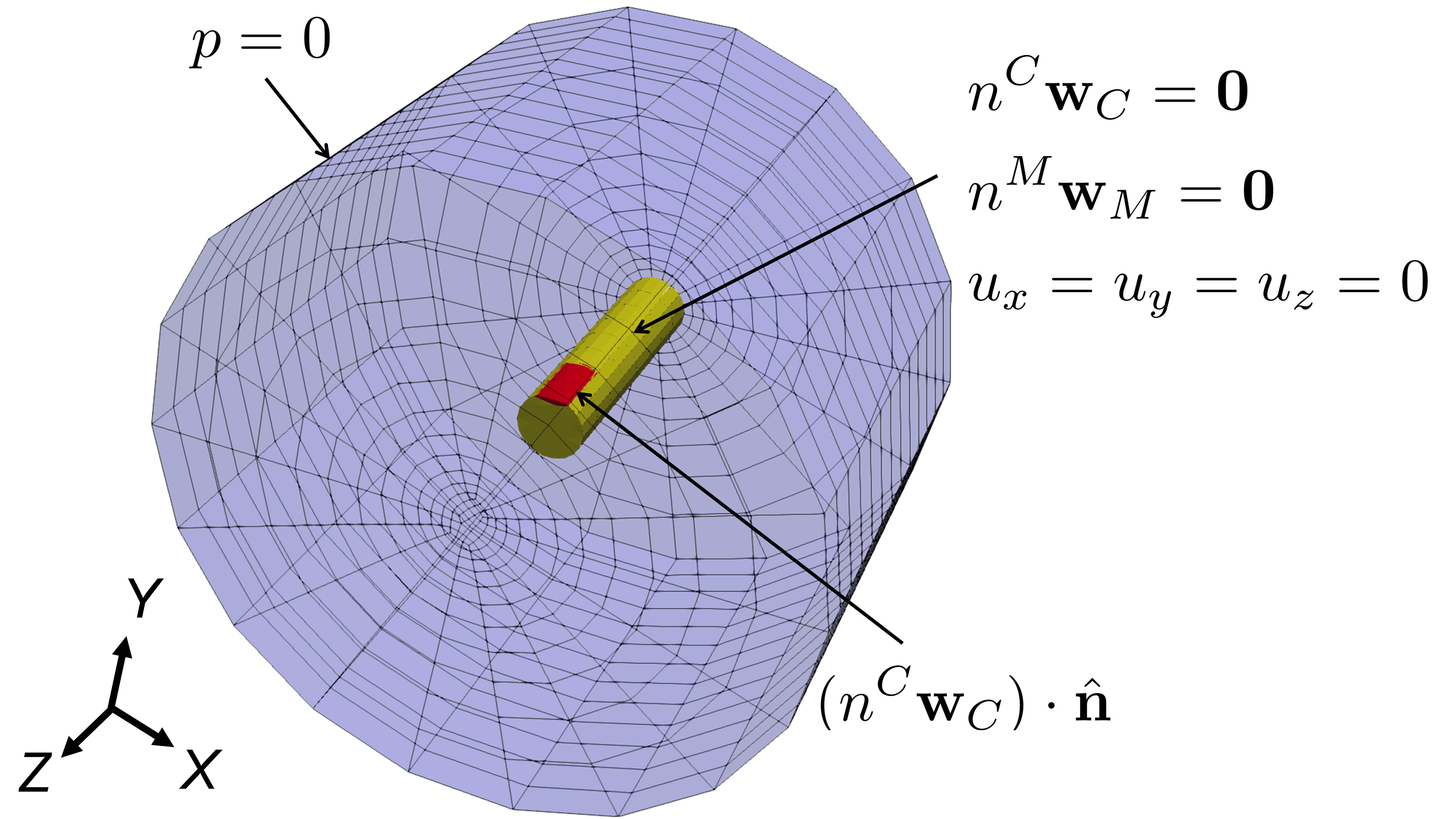}
		\caption{}
		\label{fig:boundary_conditions}
	\end{subfigure}
	\caption{\textbf{(a)} Bone cement injected into aluminium foam (above) and its schematic with dimensions in mm (below). The figure is not to scale. \textbf{(b)} Custom-made bone cement injector used for flow-controlled injection \textbf{(c)} Setup for the benchmark experiment \textbf{(d)} Geometry and boundary conditions for numerical implementation of injection inside the aluminium foam, where the cannula walls (yellow) were assigned no-slip and clamped boundary conditions, while the outer walls (blue) of the foam were assigned zero pressure. The opening for the inflow (red) was assigned velocity by dividing the flow rate by the area of the opening}
\end{figure*}

A simple problem, based on the actual vertebroplasty procedure, was formulated as a benchmark to validate the numerical model. This is described with a simplified schematic in Figure \ref{fig:schematic}. In the problem, 2 ml of bone cement is injected into a piece of aluminium foam using a surgical syringe and cannula at two different flow rates, viz., 0.1 ml/s and 0.4 ml/s. Aluminium foam was chosen since its internal structure is very similar to a vertebra \citep{loeffel_vertebroplasty_2008}.

\subsubsection{Experimental setup}
\label{sec:methods_validation_exp}

To carry out this benchmark experiment, a custom-made bone cement injector was designed and manufactured, as shown in Figure \ref{fig:exp_setup_injector}. The injector consisted of a carriage which was driven by a stepper motor using a ball screw with 5 mm feed per revolution. The stepper motor had a resolution of 1.8° corresponding to 200 steps per revolution, amounting to a calculated carriage resolution of 0.025 mm. A 200 N load cell was mounted on the carriage to measure the forces applied to the plunger of the syringe. The bone cement of brand name Vertecem V+, purchased as a non-sterile batch (OSARTIS GmbH, Germany), was used for the injection. Preparing the bone cement requires mixing the polymer powder and the monomer liquid provided in the Vertecem V+ Cement Kit (DePuy Synthes). The components were directly mixed in a 10 ml syringe to easily transfer it to the 2 ml syringes of the kit for injection, using a method described later in Section \ref{sec:methods_validation_mat}. The time from the start of mixing to the start of injection was measured using a stopwatch, which was about 200-210 seconds. The bone cement was injected by the injector at a prescribed rate into the aluminium foam, as shown in Figure \ref{fig:exp_setup_ct}. The aluminium foam was cut into a cylinder with a height of 30 mm and a diameter of 40 mm, for a size similar to that of a vertebra. The aluminium foam was placed in a polymethyl methacrylate (PMMA) housing with a hole in the centre for inserting the cannula and another hole to allow air to escape. The load cell measured the total injection force required for pushing the bone cement through the syringe and the cannula and then into the aluminium foam. Tests were also done without the aluminium foam to isolate the contribution of the aluminium foam. A Computed Tomography (CT) scanner (Revolution EVO, GE Medical Systems (Schweiz) AG, Glattbrugg, Switzerland) recorded the bone cement flow inside the foam by capturing CT images with a frequency of 2.5 Hz and a resolution of 200 x 200 x 625 microns. The CT-scanner was set at 120 kV and 150 mA, with 0.625 mm slice thickness, bone reconstruction kernel, and 0.4 s rotation time.

\subsubsection{Numerical implementation}
\label{sec:methods_validation_num}
The flow into the aluminium foam was simulated using our numerical model, the geometry and boundary conditions for which are shown in Figure \ref{fig:boundary_conditions}. The geometry and mesh were created in the software Cubit v13.0, as per the dimensions shown in Figure \ref{fig:schematic}. The geometry was discretized by a mesh of 2540 hexahedral elements. 

\subsubsection{Material parameters}
\label{sec:methods_validation_mat}

\begin{table*}[h]
\begin{center}
\begin{minipage}{\textwidth}
\begin{tabular*}{\textwidth}{@{\extracolsep{\fill}}lclcl@{\extracolsep{\fill}}}
\toprule
\textbf{Parameter}  			&  \multicolumn{2}{@{}c@{}}{\textbf{Simulation of benchmark}} &  \multicolumn{2}{@{}c@{}}{\textbf{Simulations with bone and marrow}}  \\\cmidrule{2-3}\cmidrule{4-5}%
\textbf{(Unit)}  				& \textbf{Value} & \textbf{Source} & \textbf{Value} & \textbf{Source} \\ 
\midrule
$\mu_{Lam\acute{e}}$ (GPa)      & 28.2 		& \multirow{2}{10em}{Properties of Aluminium 6101}		& 3.85 	& \multirow{2}{15em}{Properties of bone \citep{wu_youngs_2018}}\\ 
$\lambda_{Lam\acute{e}}$ (GPa)  & 54.7 		&  			& 5.77  & \\
\midrule
$\mu^C_0$ (Pa s)                & 1930  	& \multirow{4}{10em}{Using rheometer and injections through syringe and cannula without foam}	& 1930  & \multirow{4}{15em}{Same as in benchmark experiment} \\
$\mu^C_\infty$ (Pa s)       	& 1.93  	&  	& 1.93  & \\
$\lambda_{rh}^C$ (s)        	& 1.38 		&  & 1.38  & \\
$n^C_{rh}$ (-)                 	& 0.30 		&  	& 0.30  & \\ 
\midrule 
$\mu^{MR}$ (Pa s)                 	& 1.8$\times 10^{-5}$  & \multirow{4}{10em}{Viscosity of air (at 25 °C)} & - & \multirow{5}{15em}{\cite{gurkan_mechanical_2008, davis_nonlinear_2006,jansen_mechanics_2015,metzger_rheological_2014}}  \\
$\muz^M$   (Pa s)               & - 		&  		& 1000 - 0.1  & \\ 
$\muinf^M$ (Pa s)               & - 		& 	& 100 - 0.01  & \\
$\lambda^M\rh$ (s)             	& - 		& 				& 10 & \\ 
$n^M\rh$ (-)                  	& - 		& 				& 0.5 &\\
\midrule     
$\lambda_{bc}$ (-)            	& 3			& 	\multirow{4}{10em}{$\mu$CT imaging and pore-network modelling}			& 3 & \multirow{4}{15em}{Same as in benchmark experiment} \\
$n^F$ (-)                     	& 0.92 		&   & 0.92 & \\
$K^S$  (m$^2$)            		& 2.1$\times 10^{-9}$  & & 2.1$\times 10^{-9}$  &  \\
$R_{char}$ (mm)            		& 0.32 		&  		& 0.32 & \\
\botrule
\end{tabular*}
\caption{Material parameters and values used for numerical implementation of the benchmark problem and simulations with marrow}
\label{tab:mats_benchmark}
\end{minipage}
\end{center}
\end{table*}

\begin{figure*}[t]
	\begin{subfigure}[t]{0.49\textwidth}
		\centering
		\includegraphics[width=\textwidth]{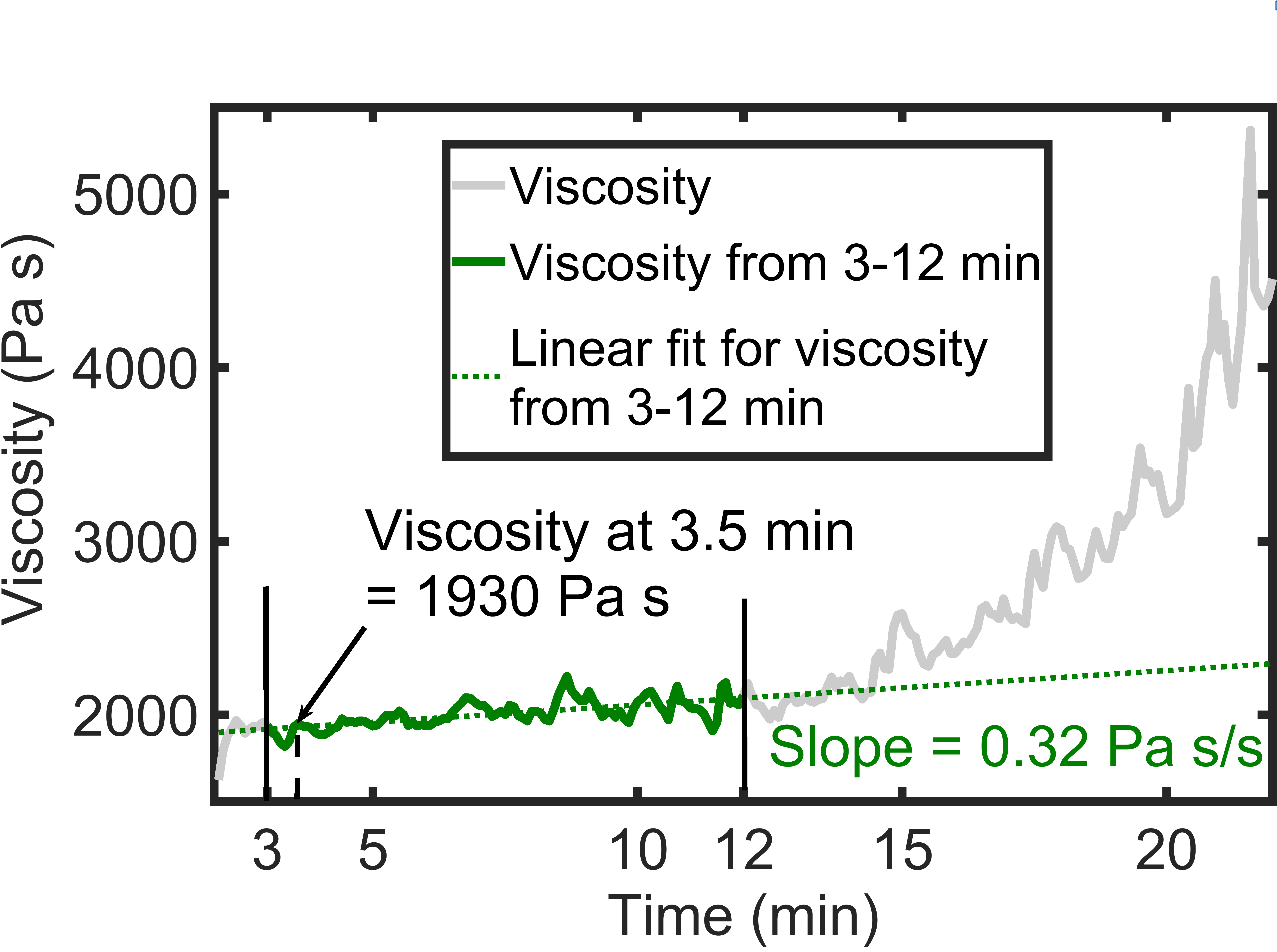}
		\caption{}
		\label{fig:rheo_parameter_left}
	\end{subfigure}
	\begin{subfigure}[t]{0.49\textwidth}
		\centering
		\includegraphics[width=\textwidth]{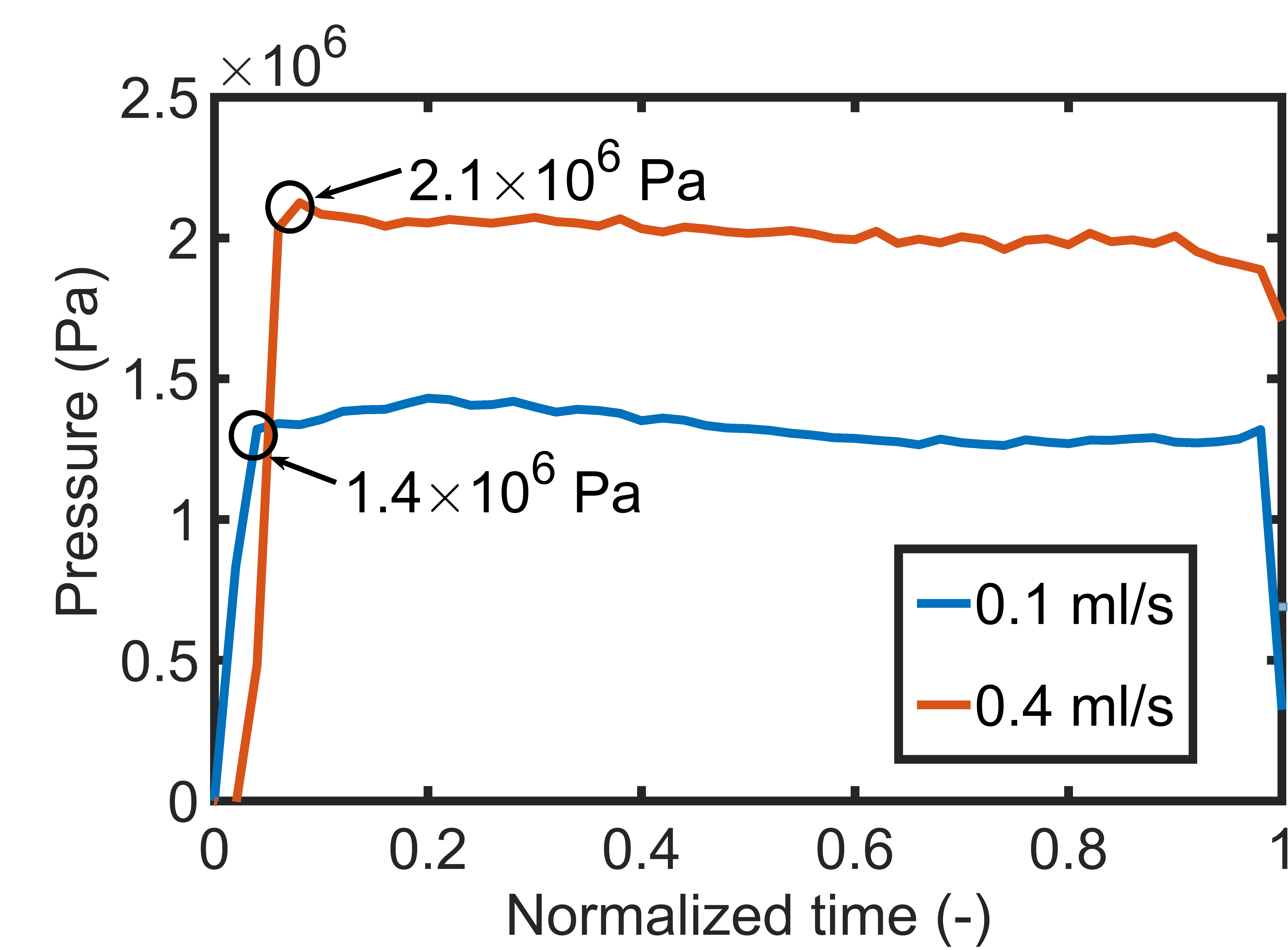}
		\caption{}
		\label{fig:rheo_parameter_right}
	\end{subfigure}
	\begin{subfigure}[b]{\textwidth}
		\centering
		\includegraphics[width=\textwidth]{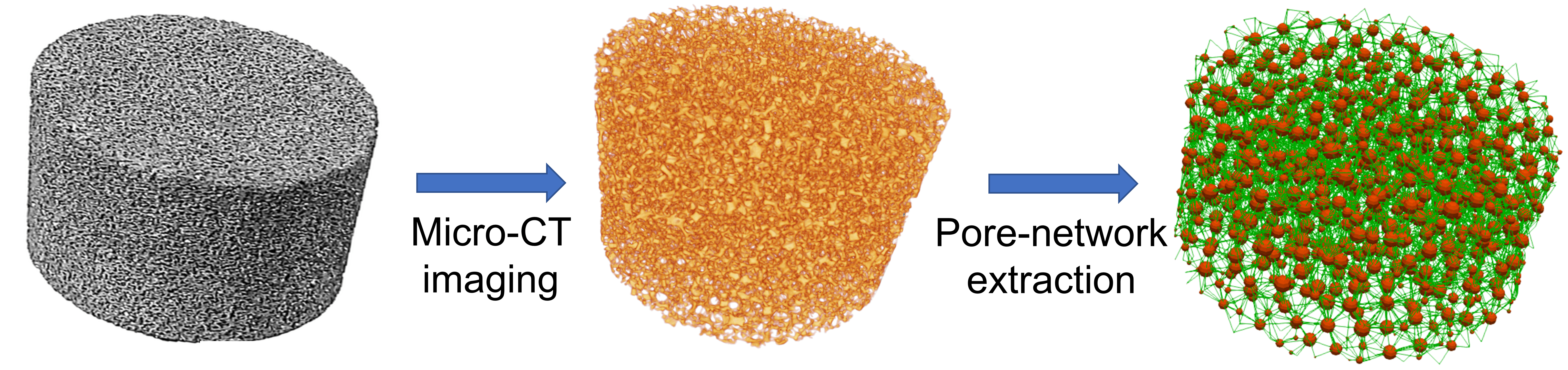}
		\caption{}
		\label{fig:pnm}
	\end{subfigure}
	\caption{\textbf{(a)} Measurements used for obtaining rheological parameters of bone cement: change in viscosity with time due to curing (left) and injection pressure required for cement flow through the injection equipment without the aluminium foam (right) \textbf{(b)} Pore-network extraction from micro-CT image of the aluminium foam}
\end{figure*}

The parameters and their values used for the benchmark problem are summarized in Table \ref{tab:mats_benchmark}. The aluminium foam was made of standard material Aluminium 6101, thus the material parameters were easily available from literature. Since the aluminium foam consisted of nearly uniformly sized pores, $\lambda_{bc} = 3$ was taken as a reasonable estimate.

The rheological parameters of the bone cement include the upper and lower viscosity limits $\muz^C$ and $\muinf^C$, the relaxation time $\lambda^C\rh$ for the transition to shear-thinning behaviour, and the flow behaviour index $n^C\rh$ for the shear-thinning. The upper viscosity limit and its change with time were determined by rheological measurement. We used multiple methods of preparation of bone cement for rheological measurements to simulate real kit application. After multiple trials, the successful and reproducible method consisted of the following steps:
\begin{enumerate}
	\item Pre-weighing 2.6 grams of the polymer powder in a 10 ml beaker.
	\item Separately, 10 ml of the monomer liquid was prepared in a batch glass vial.
	\item At time t = 0 s, 1.0 ml monomer liquid was dropped into the beaker with the polymer powder using a positive displacement pipette. 
	\item Immediately at that time (t = 0 s), a timer was started to count 20 seconds of gentle stirring using a polyetheretherketone stirring rod (counting ~20 rotations). 
	\item After 20 s, part of the sample was gently dropped onto the rheometer bottom plate and the top plate was gently lowered down. The humidity control hood was used to avoid sample evaporation, and silicone oil was gently spread around the sample directly before each measurement.  
\end{enumerate}
Understanding the timing of polymerization is crucial; therefore, we evaluated the average time taken from the start of the mixing of cement components to the start of the rheological measurement with the above-established procedure, which was 118 ± 10 s (based on a total of 26 trials). The rheological measurements were performed on MCR302 rheometer from Anton Paar. Parallel plate geometry (PP) with a 25 mm diameter top plate and a 1.5 mm gap was used. Time sweep measurement using plate-plate with a total measurement time of 20 minutes was run in a rotational mode at a shear rate of 0.2 s$^{-1}$. Each point was recorded every 5 seconds. The results are shown in Figure \ref{fig:rheo_parameter_left}. The viscosity at 3.5 minutes, i.e.~the cement preparation time in the injection experiments, was used to obtain $\muz^C=1930$ Pa s. The actual value of the lower viscosity limit $\muinf^C$ is difficult to obtain due to its occurrence at shear rates too high to be able to measure reliably, although the viscosity has been observed to reduce by at least two orders of magnitude without plateauing \citep{krause_viscosity_1982, lepoutre_bone_2019}. Therefore, the value $\muinf^C=1.93$ Pa s was assumed such that a 1000-times reduction in viscosity is allowed. The cement showed a sharp increase in viscosity for about first 2-2.5 minutes after the start of mixing, after which the curing slowed down for the next 9-10 minutes, followed by rapid curing at the end. In clinical applications and our benchmark experiment, the injection occurs in the period from 3 to 12 minutes, where the viscosity increase was found to be linear at 0.32 Pa s/s. This increase would have a negligible effect on the viscosity during the period of injection for our experiments. Hence, the dependence on time was neglected. 

To obtain the remaining two parameters, $\lambda^C\rh$ and  $n^C\rh$, the experimental setup as explained in Section \ref{sec:methods_validation_exp} was used without the aluminium foam to obtain the injection force required for flow through the assembly made of the syringe, the nozzle, and the cannula; at flow rates 0.1 and 0.4 ml/s. The injection pressure was then obtained by dividing the force by the area of the syringe. The injection pressures are shown in Figure \ref{fig:rheo_parameter_right}. The value immediately after the initial ramp is taken. These injection pressures and their respective flow rates were inserted in the analytical solution for the pressure required by a Carrreau fluid flowing through a tube at a given flow rate, as given in \citep{sochi_analytical_2015}. This gave a system of two equations, which could be solved to obtain the two remaining unknowns $\lambda^C\rh = 1.38$ and $n^C\rh = 0.30$. All parameters are listed in Table \ref{tab:mats_benchmark}.

The remaining parameters, viz., porosity, permeability, and the characteristic radius $R_{char}$ (used in the average viscosity rheological model), are dependent on the pore geometry of the porous medium. A micro-CT scan of the aluminium foam was carried out to resolve the pore structure of the aluminium foam, from which the porosity could be computed. The micro-CT scanner used here was a VivaCT40 from SCANCO Medical AG at an isotropic resolution of 19 µm using 70 kV, 114 µA and 200 ms integration time. To determine the permeability and the characteristic radius, we used a pore-network model \citep{joekar-niasar_non-equilibrium_2010}. The micro-CT image was used to extract a pore-network, as shown in Figure \ref{fig:pnm}, using an open-source Python toolkit PoreSpy \citep{gostick_porespy_2019}. The permeability was obtained from this pore-network by simulating a Stokes flow through this pore-network using the open-source porous media flow solver DuMux \citep{koch_dumux_2020}. Note that the permeability obtained here was isotropic, i.e., nearly the same in all directions, hence we could use a scalar value instead of the tensor in Equations \ref{eq:darcyc} and \ref{eq:darcym}. Similarly, the characteristic radius $R_{char}$ was obtained from the volume-weighted average of the radii of the pores and the throats.

\subsection{Clinically relevant simulations and further investigations}
\label{sec:further_trials}

To investigate a more clinically relevant setting, we used the same geometry and boundary conditions as in the benchmark problem with Cannella model and 0.4 ml/s flow rate, while suitably replacing the values of material parameters. Firstly, we replaced the mechanical properties of the aluminium foam with those of trabecular bone \citep{wu_youngs_2018}. Since the mechanical properties of the trabeculae vary depending on factors like specimen, location, and condition; mean values found in the literature were used. Furthermore, instead of a previously empty porous medium, we did trials assuming the presence of bone marrow. In reality, bone marrow could consist of red bone marrow, which due to its similarity to blood, is non-Newtonian by nature; and yellow bone marrow, which has Newtonian rheology \citep{gurkan_mechanical_2008}. The rheological behaviour could vary depending on the red-to-yellow bone marrow ratio, which is then dependent upon the age and health conditions of the patient. Given this uncertainty, we did the simulations over a range of viscosities, considering both non-Newtonian and Newtonian rheologies. For non-Newtonian rheologies, we assumed realistic values for the parameters based on literature \citep{gurkan_mechanical_2008, davis_nonlinear_2006, metzger_rheological_2014, jansen_mechanics_2015}, while the viscosity limits were set such that $\muz^M/\muinf^M = 10$. The rest of the parameters used were the same as in the benchmark experiment. The parameters and their values are summarized in Table \ref{tab:mats_benchmark}.

Apart from the above, further trials were carried out to investigate the effect of permeability, porosity, and the Brooks-Corey uniformity parameter $\lambda_{bc}$. In reality, these parameters are interdependent, e.g., a higher porosity also leads to a higher permeability. However, here we wanted to investigate them in isolation to study their standalone effects. The chosen values for permeability and porosity were those obtained from literature for human vertebrae \citep{baroud_experimental_2004, daish_estimation_2017, ochia_hydraulic_2002}. As far as the author is aware, no information is available in the literature for the Brooks-Corey uniformity parameter of human vertebrae. Therefore, a general but wide range of values from 0-10 was chosen. 

\section{Results}
\label{sec:results}

\subsection{Validation with benchmark problem}

\begin{figure*}[htbp]
	\centering
	\begin{subfigure}[t]{0.9\textwidth}
		\centering
		\includegraphics[width=\textwidth]{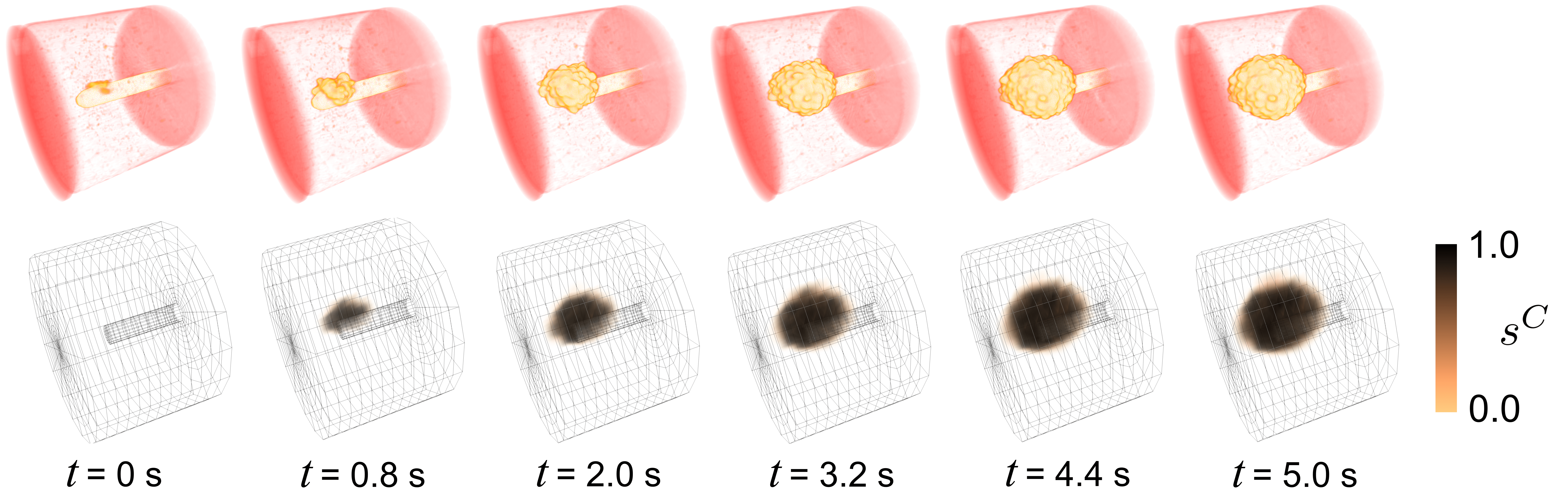}
		\caption{}
		\label{fig:injexp_timelapse}
	\end{subfigure}
	
	\begin{subfigure}[b]{\textwidth}
		\centering
		\includegraphics[width=\textwidth]{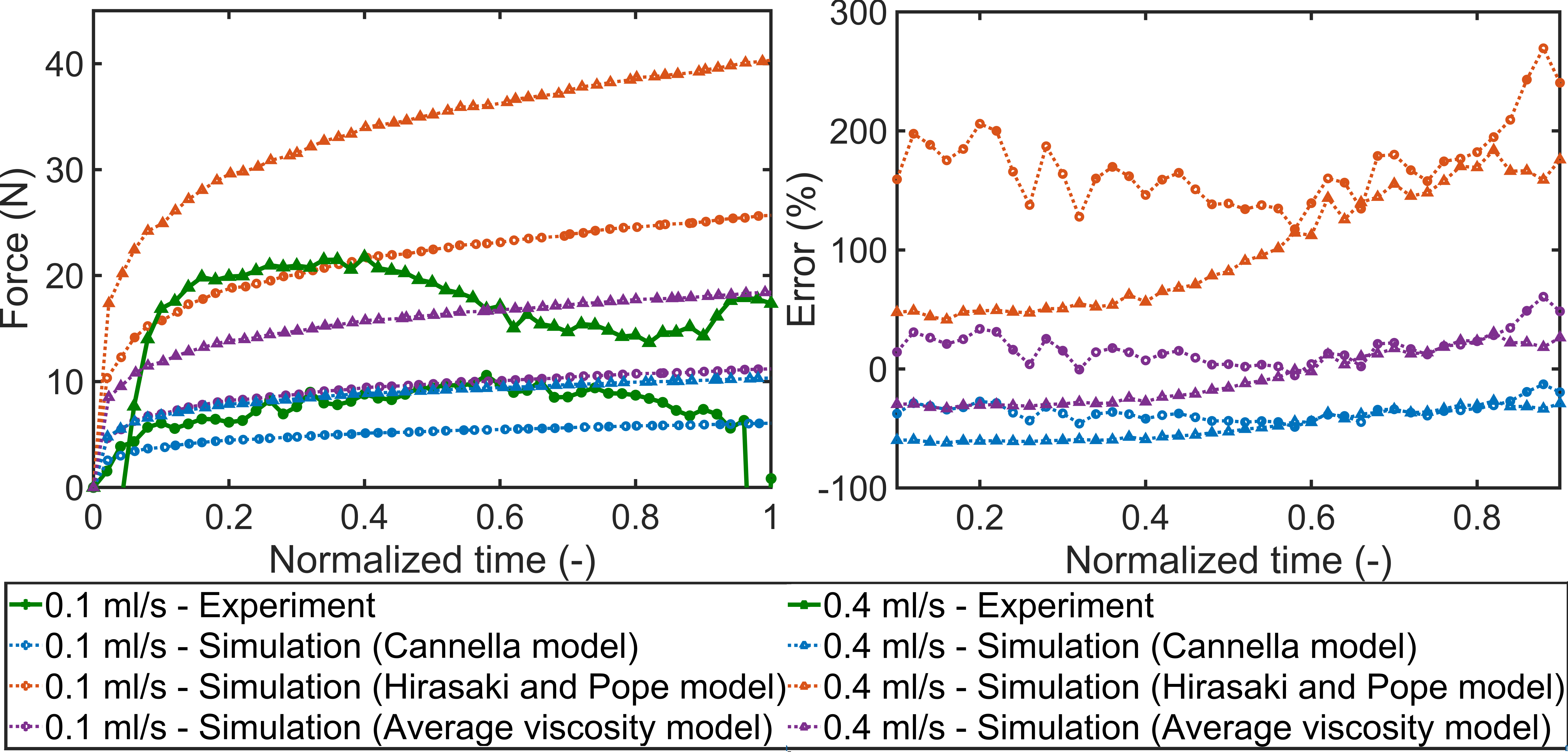}
		\caption{}
		\label{fig:validation_injpres}
	\end{subfigure}
	
	\caption{\textbf{(a)} Cement flow inside the aluminium foam at various time steps recorded during the benchmark experiment using CT images (top) and cement saturation ($s^C$) as obtained from simulation (bottom)  for 0.4 ml/s flow rate \textbf{(b)} Injection force measured from experiments compared to simulations using various rheology upscaling models (left) and their respective percentage errors (right)}
	\label{fig:results_benchmark}
\end{figure*}

Looking at the cement distribution first in Figure \ref{fig:injexp_timelapse}, similar cement distribution patterns, i.e., fully saturated ($s^C \approx 1$) with a sharp front, were obtained from the simulation and in the experiment of the benchmark problem. However, the injection forces, shown in Figure \ref{fig:validation_injpres}, differed from the experiments depending on which rheology upscaling model was used. The difference also did not remain constant with time. Table \ref{tab:perc_error} shows the percentage error averaged over the time, barring the 10\% at the start and the end to ignore any effect of the difference in climb/drop times. The average viscosity model gave results closest to the experiments. The Cannella model gave results with more error but in a similar range whereas the Hirasaki model performed the worst with much higher magnitudes of force. 

\begin{table}[h]
	\begin{center}
		\begin{minipage}{174pt}
			\begin{tabular}{@{}lcc@{}} 
				\toprule
				\multirow{2}{10em}{\textbf{Rheology upscaling model}} & \multicolumn{2}{@{}c@{}}{\textbf{Error (RMS)}} \\ \cmidrule{2-3}
				\textbf{} & \textbf{0.1 ml/s} & \textbf{0.4 ml/s} \\ 
				\midrule
				Cannella model & 37\% & 50\% \\
				Hirasaki and Pope model & 173\% & 110\% \\
				Average viscosity model & 22\% & 23\% \\ 
				\botrule
			\end{tabular}
			\caption{Root mean square error (RMS) for rheology upscaling models over 0.1 to 0.9 of normalized time}
			\label{tab:perc_error}
		\end{minipage}
	\end{center}
\end{table}

\subsection{Clinically relevant simulations}

\begin{figure*}[htbp]
	\begin{subfigure}[t]{0.49\textwidth}
		\centering
		\includegraphics[width=\textwidth]{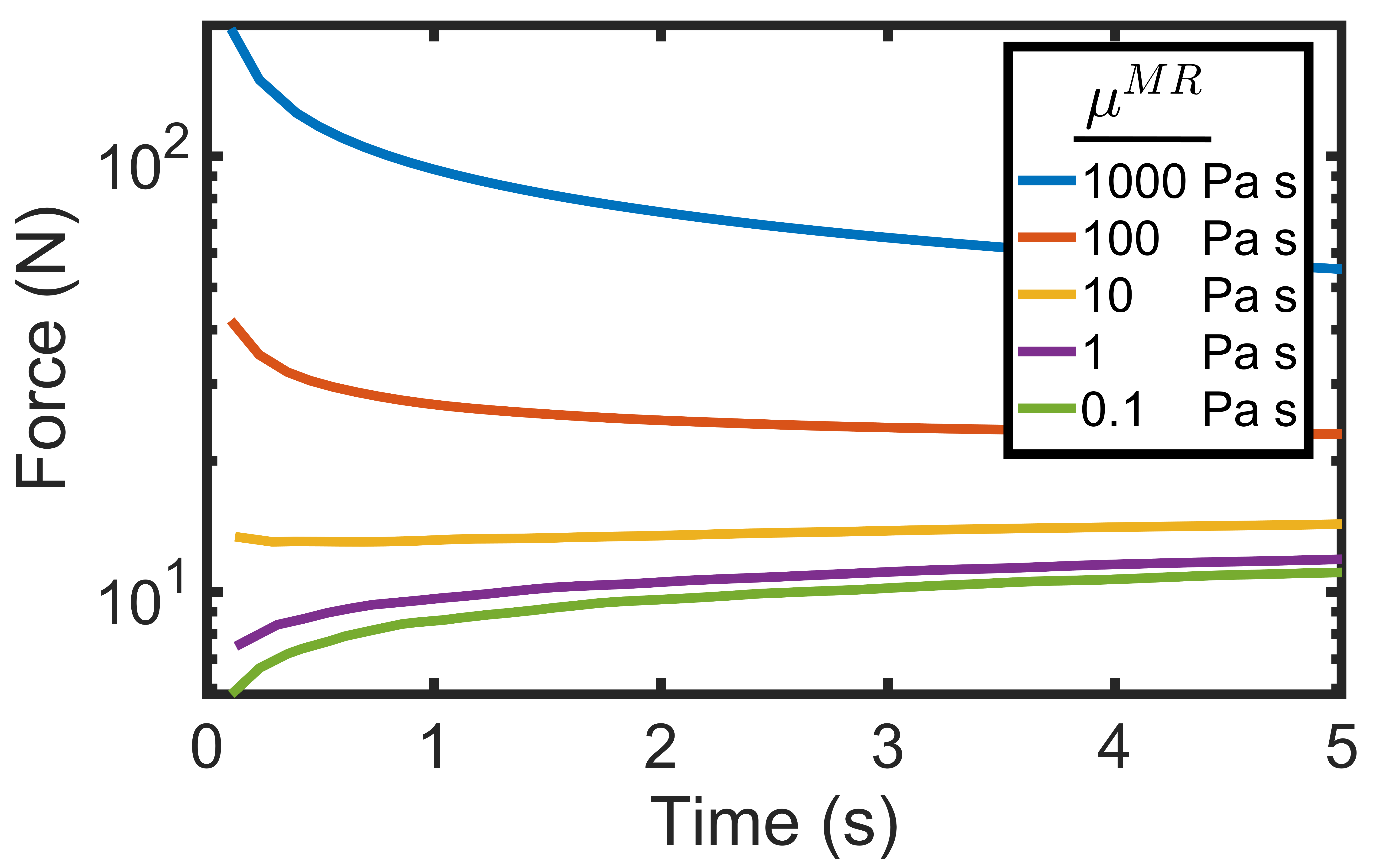}
		\caption{}
		\label{fig:marrow_injpres_n}   
	\end{subfigure}
	\begin{subfigure}[t]{0.49\textwidth}
		\centering
		\includegraphics[width=\textwidth]{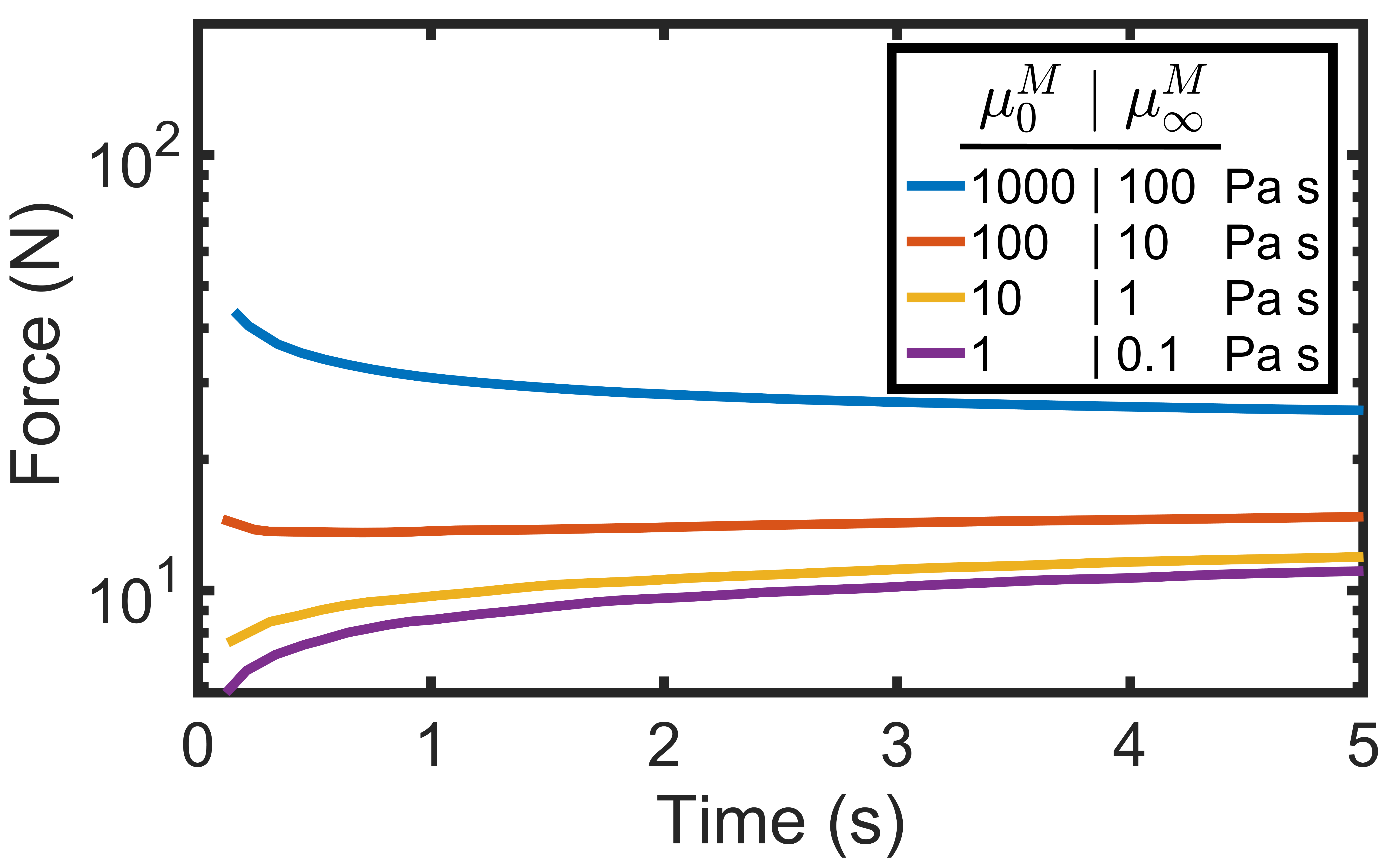}
		\caption{}
		\label{fig:marrow_injpres_nn}   
	\end{subfigure}
	\begin{subfigure}[b]{0.5\textwidth}
		\centering
		\includegraphics[width=\textwidth]{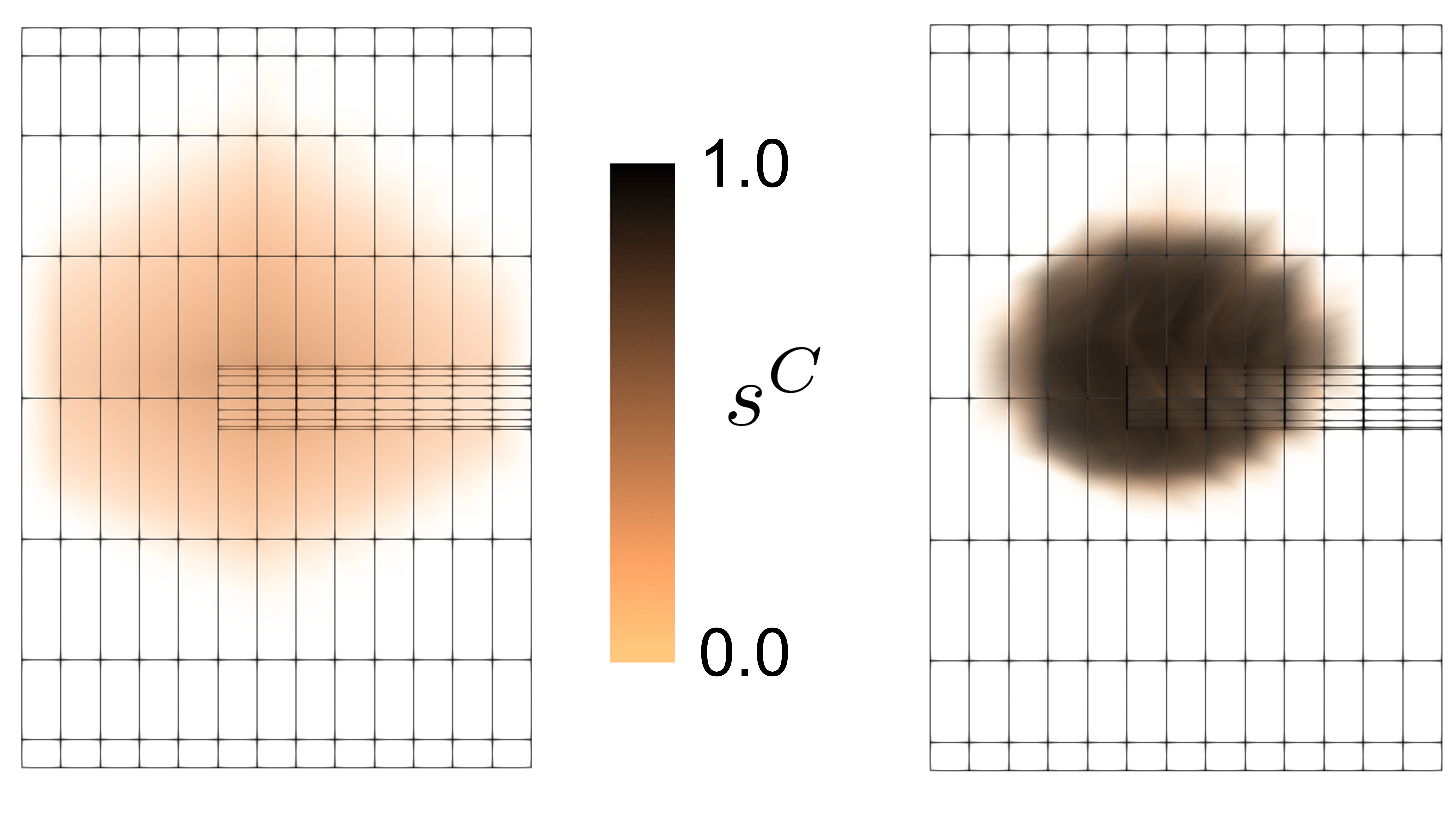}
		\caption{}
		\label{fig:cement_dist_compare}   
	\end{subfigure}
	\hfill
	\begin{subfigure}[b]{0.42\textwidth}
		\centering
		\includegraphics[width=0.9\textwidth]{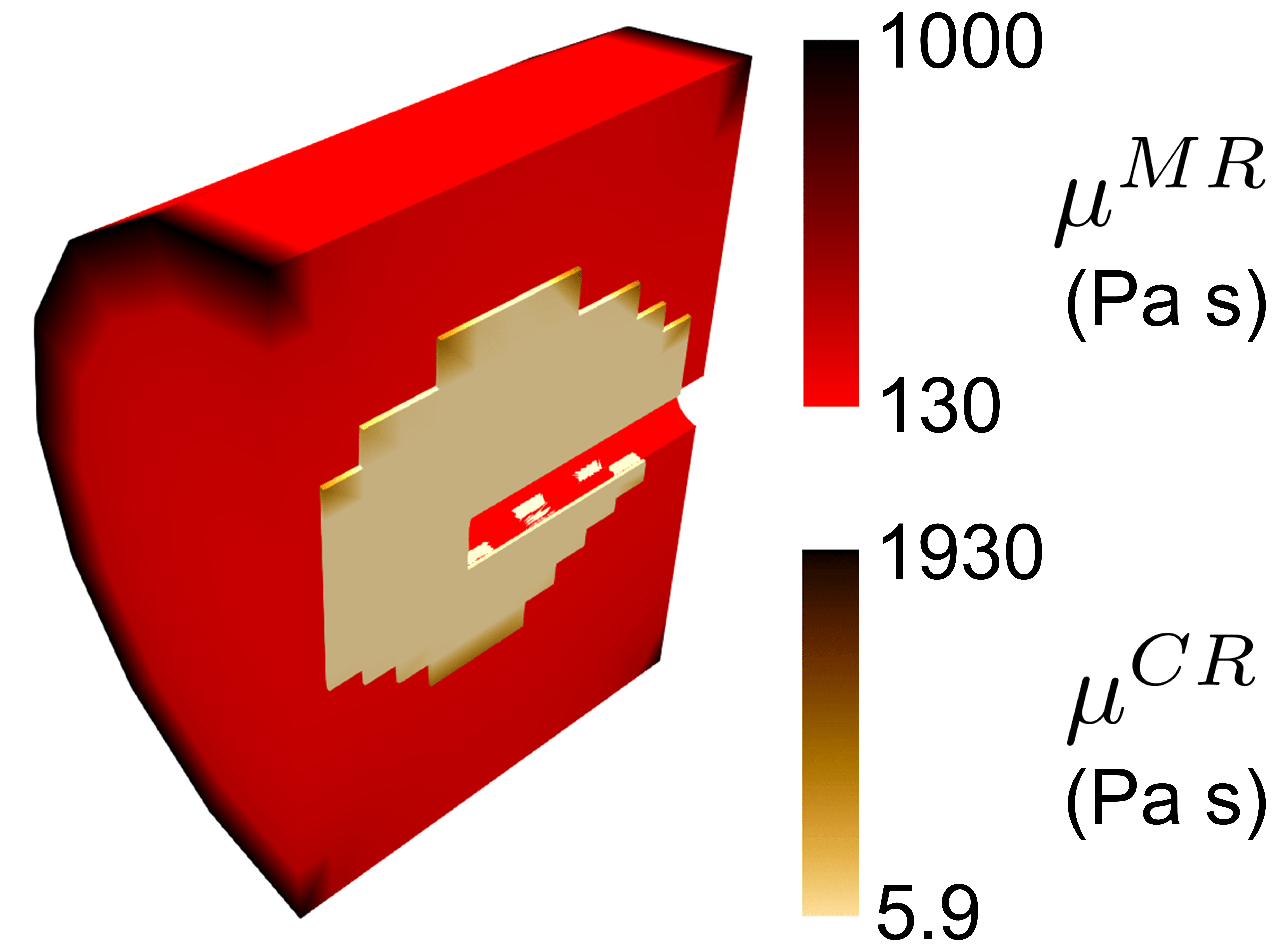}
		\caption{}
		\label{fig:marrow_cement_visc}   
	\end{subfigure}
	\caption{Injection force required to inject cement inside a marrow-saturated foam/vertebra for \textbf{(a)}  various Newtonian marrow viscosities and \textbf{(b)} various non-Newtonian marrow viscosities \textbf{(c)} Cement saturation ($s^C$) inside the marrow-saturated foam/vertebra for the cases of Newtonian bone marrow with viscosity $\mu^{MR}=1000$ Pa s (left) and $\mu^{MR}=0.01$ Pa s (right) \textbf{(d)} Cut section showing viscosity distribution of cement ($\mu^{MR}$) and marrow ($\mu^{CR}$) marrow-saturated foam/vertebra for the case of $\muz^M = 10^3$, $\muz^M = 10^2$ Pa s}
\end{figure*}

From Figure \ref{fig:marrow_cement_visc}, it was observed that except at the peripheries, the viscosities were the same everywhere within the fluids despite the non-Newtonian rheology. The cement viscosity after shear-thinning was 5.9 Pa s. Figure \ref{fig:marrow_injpres_n} shows a stark difference in the injection force development in cases with marrow viscosity higher than this value of 5.9 Pa s, compared to the other way round. For cases where marrow viscosity was higher, the force peaked near the start and declined afterwards; whereas when marrow viscosity was lower, the force showed a small gradual increase with time. The difference in the injection force between the non-Newtonian and the Newtonian marrow cases was only in magnitude, with higher values observed in the case of Newtonian marrow since it does not undergo shear-thinning. Another difference caused due to the marrow viscosity is shown in Figure \ref{fig:cement_dist_compare}. Higher marrow viscosity caused the cement distribution to be diffuse, as observed from low cement saturation values; and to spread farther, nearly touching the end of the porous medium. For lower marrow viscosity, the cement cloud was fully saturated and compact, similar to the benchmark experiment. Note that with both the aluminium foam and the trabecular bone as the solid skeleton, the negligible deformations in the order of 10$^{-10}$ m and 10$^{-9}$ m respectively were obtained. The results can be referred to in the dataset \cite{darus-3146_2022}.

\subsection{Permeability, porosity, and Brooks-Corey uniformity parameter}

\begin{figure*}[htbp]
	\begin{subfigure}[t]{\textwidth}
		\centering
		\includegraphics[width=\textwidth]{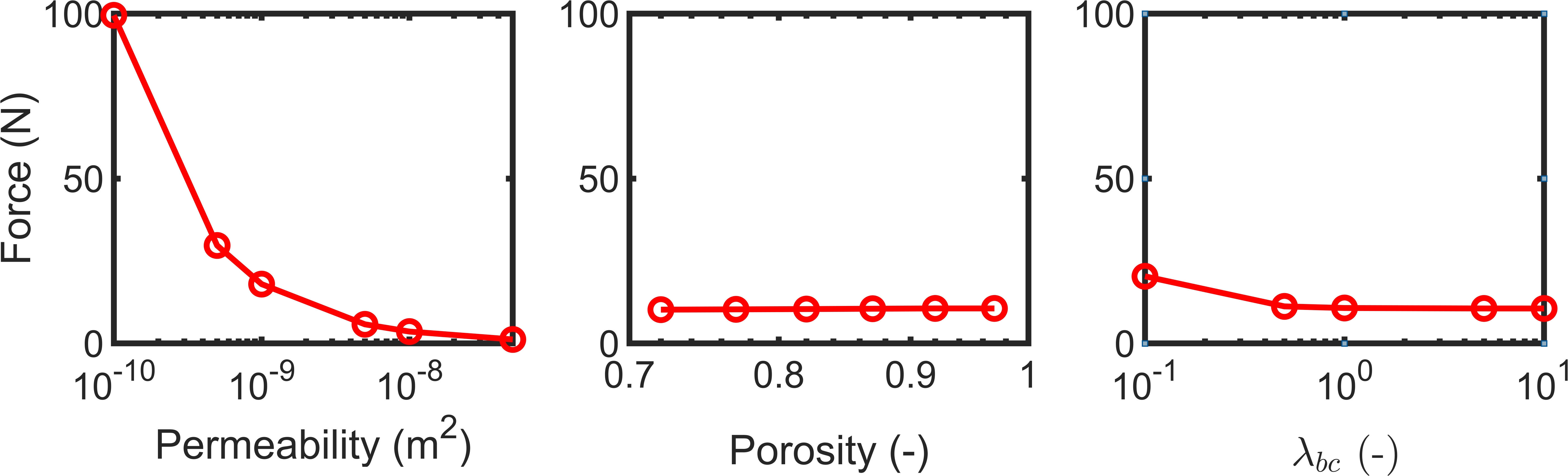}
		\caption{}	
		\label{fig:permporobc}
	\end{subfigure}
	\begin{subfigure}[b]{\textwidth}
		\centering
		\includegraphics[width=\textwidth]{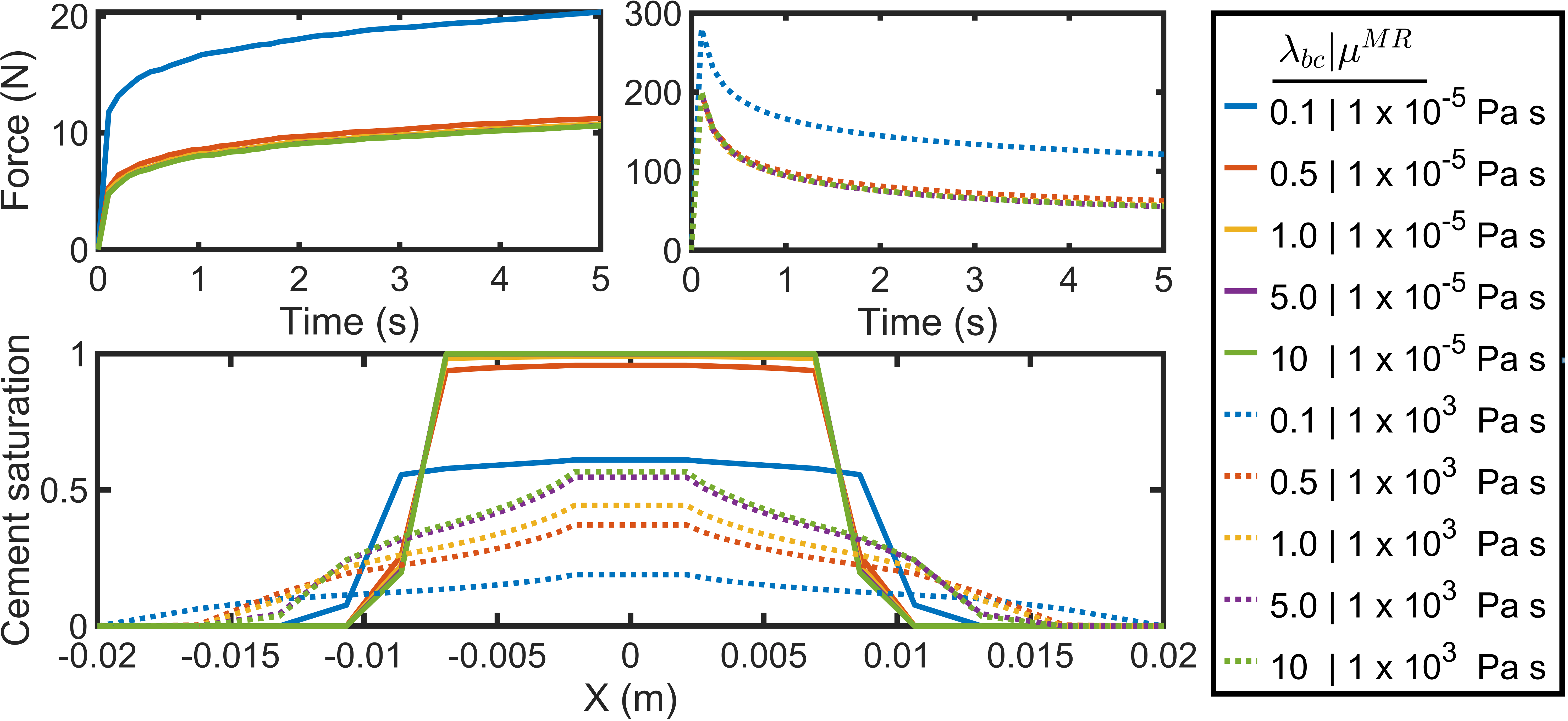}
		\caption{}
		\label{fig:bc_analysis}
	\end{subfigure}	
	\caption{\textbf{(a)} Variation in the injection force (at end time step of the simulation) with  permeability, porosity, and $\lambda_{bc}$ \textbf{(b)} The injection force development and the distribution of cement saturation (along X-axis at the point of injection) for various values of $\lambda_{bc}$ when marrow has higher viscosity (dotted lines) and lower viscosity (solid lines) than the cement}
	\label{fig:more_params}
\end{figure*}

Figure \ref{fig:more_params} shows how the injection force at the end of injection varies with permeability, porosity, and the uniformity parameter $\lambda_{bc}$. Only permeability showed a significant influence out of the three parameters. Porosity had nearly no influence. For $\lambda_{bc}$, there was some increase in injection force at $\lambda_{bc}=0.1$, but that is already an extreme and unrealistic case. However, since $\lambda_{bc}$ represents the pore size uniformity in the porous medium, we also investigated its effect on the cement distribution. Therefore, we considered two cases: with lower marrow viscosity (10$^{-5}$ Pa s) and with higher marrow viscosity (10$^{3}$ Pa s) relative to the cement viscosity since we observed a difference in cement distribution for these cases earlier in Figure \ref{fig:cement_dist_compare}. Accordingly, we compared the injection force graphs and cement distributions for various values of $\lambda_{bc}$ for both these cases, as shown in Figure \ref{fig:bc_analysis}. We observed no significant difference in the injection forces due to $\lambda_{bc}$ as long as $\lambda_{bc} > 0.1$. However, the cement distribution did show dependence on $\lambda_{bc}$ for the case of higher marrow viscosity, although this was not the case when the marrow viscosity was lower than the cement. The exception here is again the extreme case of $\lambda_{bc} = 0.1$. 

\section{Discussion}
\label{sec:discussion}

The matching cement patterns in the experiment and the simulations for the benchmark problem show that the presented model can correctly describe and replicate the flow boundary conditions from the experiments. In terms of injection force, the average viscosity model for upscaling the rheologies gave the best results, followed by the Cannella model. The error, although not negligible, is expected due to uncertainty caused by factors like the sensitivity of the cement curing to the mixing conditions and exposure to air, which are difficult to control given the constraints of the rheological measurement and the benchmark experiments. The Hirasaki and Pope model performed the worst among the three. In contrast, the study by \cite{eberhard_determination_2019} found the results from Hirasaki and Pope and the average viscosity models similar and more accurate than the Cannella model. The reason for better accuracy of the Hirasaki and Pope model in their study could be that they used a ``granular" structure made of monodisperse spheres, which is also what \cite{hirasaki_analysis_1974} used to arrive at the value of $\mathbb{C} = 0.69$ (in Equation \ref{eq:rheo_upscaling}) in their model. Comparatively, the ``fibrous"-type structure of the aluminium foam is considerably different. Therefore, the Cannella and the average viscosity models are better suited than the Hirasaki and Pope model to upscale the viscosity for the case of aluminium foam, and therefore, also for vertebrae. The average viscosity model gave the most accurate results not only in the work of \cite{eberhard_determination_2019}, but also in ours, despite the difference in the porous materials used, hinting at its potential applicability to a relatively wide range of porous media. The reason for its wider applicability compared to the semi-empirical models could be because it requires an additional parameter (the characteristic radius $R_{char}$) derived from the pore geometry, whereas the other two models use fixed values for the constant $\mathbb{C}$. Interestingly, the injection force increased with time in the simulations even though the time-dependent increase of viscosity due to curing was neglected in the model. Therefore, the increase in the force must have to do with the nature of the cement distribution. Also, note that in both experiment and simulation, a four-times increase in flow rate requires less than double the injection force. The non-linear dependence is due to the decrease in the cement viscosity by shear-thinning at higher flow rates.

The presence of bone marrow had a considerable impact on the results. Whether the marrow is non-Newtonian or Newtonian affects only the magnitude of the injection force, the lesser force being in the non-Newtonian case due to the shear-thinning viscosity. The injection force development and cement distribution are significantly affected by whether the cement viscosity is higher or lower than the marrow viscosity. There are three clear risks if the cement viscosity falls below that of the marrow: (i) unintuitive injection force development, (ii) improper cement filling, and (iii) additional dependence on the pore size uniformity. Firstly, the required injection force is highest at the start, followed by a substantial dip. The peak force needed at the start requires the practitioners to apply more effort early on, which could lead to sudden excess cement injection as the required force suddenly dips. Hence, such injection force behaviour is highly unintuitive for practitioners relying on haptic feedback. Secondly, the injected cement only partially fills the porous bone before spreading further. The marrow is not fully displaced because its higher viscosity makes it less movable than the cement. Therefore, the cement spreads in an unstable fingering pattern in a phenomenon known as ``viscous fingering". Such a filling pattern causes poor interdigitation of the cement with the trabeculae. It also causes the cement to spread farther, potentially leaking outside the cortical shell before the required volume is injected. Finally, since the fingers develop depending on the pore geometries, an additional dependence of the cement distribution on the pore size uniformity parameter $\lambda_{bc}$ is introduced. This additional dependence adds another factor of uncertainty to the procedure outcome. On the other hand, these risks could be avoided if the cement viscosity stays higher than the marrow's, yielding advantages like (i) a gradually rising injection force, which is more stable and intuitive for practitioners; (ii) the cement completely filling the porous bone, providing better interdigitation and leading to better mechanical strengthening; (iii) no strong dependence on the pore size uniformity. 

The advantages of high-viscosity cement have been emphasized in earlier works \citep{baroud_high-viscosity_2006, loeffel_vertebroplasty_2008}, but none have provided a clear definition for `high-' or `low-viscosity' cement. From the results of this work, we can state that the marrow viscosity is the reference relative to which the cement viscosity must be higher, especially after the shear-thinning from the injection. The shear-thinning causes the cement and marrow viscosities to come closer to their respective $\muinf$ compared to $\muz$. In this regard, the power law exponent $n$ and the lower viscosity limit $\muinf$ of the cement and the marrow are crucial. For practitioners, this information could be useful for choosing the bone cement and its curing time depending on the patient's marrow composition, or for making decisions like whether techniques like marrow aspiration need to be employed. 

There are also some limitations to our current study. As mentioned earlier, the bone cement viscosity and curing behaviour are sensitive to mixing conditions and exposure to air. We also did not consider the temperature dependence, which would cause the viscosities to change from the heat released by curing, or the difference in the room and body temperature in a clinical setting. More limitations could arise in clinical settings, e.g., we applied boundary conditions such that the fluids can freely flow out of the walls, whereas \textit{in vivo}, they might enter into the blood vessels inside the vertebra. Modelling these \textit{in vivo} conditions would require dependence of the boundary conditions on the distribution of the blood vessels in the vertebra. Moreover, influential parameters like permeability and marrow viscosity are hard to determine in \textit{in vivo} conditions. These parameters could be estimated from \textit{in vivo} measurements like clinical CT images \citep{teo_correlation_2007} or patient characteristics based on clinical data. However, these factors contribute to the uncertainty in the outcome of the simulation results. In this regard, a study for quantifying the uncertainties in the parameters could be quite beneficial. Simulations using such estimations could help practitioners estimate at least a pilot range for the operating parameters, e.g.~injection pressure, cement viscosity, curing time, flow rate, etc., before performing the actual procedure instead of completely relying on haptic feedback and frequent X-ray images at the time of the procedure. The reduced reliance on X-ray imaging would decrease the patient's exposure to harmful radiation. The model could also serve as a training tool for relatively young, inexperienced practitioners. In all the simulations carried out in this study, the deformations were negligible: $\approx 10^{-10}$ m for aluminium foam, $\approx 10^{-9}$ m for trabecular bone. Despite this, the capability to compute deformations is advantageous in cases like a weakened vertebra where the deformations may not be negligible. Extending the model to include the effect of existing fractures on vertebroplasty is possible and especially clinically relevant for future work since vertebroplasty mostly happens on fractured vertebrae. Similarly, the inclusion of temperature effects in the model has relevance since tissue necrosis is another risk associated with vertebroplasty. Lastly, the physics of this model can be suitably modified for other injection and infusion processes in the biomedical field.

\section{Conclusion}
\label{sec:conclusion}

We presented a continuum-mechanical model based on the Theory of Porous Media for simulating vertebroplasty. The model can simulate the injection of bone cement in porous materials like vertebrae and yield realistic results, as was evident from the benchmark experiment used for validation. The average viscosity model is found to be the most suitable approach for upscaling the rheology in the macro-scale framework. The Cannella model could potentially be used, but the Hirasaki and Pope model is not suitable. Our simulations show that the cement must have a higher viscosity than the marrow to ensure stable development of injection force and proper filling of cement. In this regard, the marrow's and the cement's relative rheologies play a crucial role in the outcome of vertebroplasty and in avoiding cement leakage. We expect the current model to support future developments of vertebroplasty simulations that are closer to clinical reality and expect its possibilities to be extended toward modelling other fluid injection mechanisms in biomedical fields.

\bibliography{sn-bibliography}

\begin{thebibliography}{}
\providecommand{\doi}[1]{\url{https://doi.org/#1}}
\bibcommenthead

\bibitem [\protect \citeauthoryear {%
Baroud%
, Crookshank%
\BCBL {}\ \BBA {} Bohner%
}{%
Baroud%
\ \protect \BOthers {.}}{%
{\protect \APACyear {2006}}%
}]{%
baroud_high-viscosity_2006}
\APACinsertmetastar {%
baroud_high-viscosity_2006}%
\begin{APACrefauthors}%
Baroud, G.%
, Crookshank, M.%
\BCBL {} Bohner, M.%
\end{APACrefauthors}%
\unskip\
\newblock
\APACrefYearMonthDay{2006}{}{}.
\newblock
{\BBOQ}\APACrefatitle {High-{Viscosity} {Cement} {Significantly} {Enhances}
  {Uniformity} of {Cement} {Filling} in {Vertebroplasty}: {An} {Experimental}
  {Model} and {Study} on {Cement} {Leakage}:} {High-{Viscosity} {Cement}
  {Significantly} {Enhances} {Uniformity} of {Cement} {Filling} in
  {Vertebroplasty}: {An} {Experimental} {Model} and {Study} on {Cement}
  {Leakage}:}.{\BBCQ}
\newblock
\APACjournalVolNumPages{Spine J}{31}{22}{2562--2568}.
\newblock
\begin{APACrefURL} {https://doi.org/10.1097/01.brs.0000240695.58651.62}
  \end{APACrefURL}
\newblock

\newblock

\PrintBackRefs{\CurrentBib}

\bibitem [\protect \citeauthoryear {%
Baroud%
, Falk%
, Crookshank%
, Sponagel%
\BCBL {}\ \BBA {} Steffen%
}{%
Baroud%
\ \protect \BOthers {.}}{%
{\protect \APACyear {2004}}%
}]{%
baroud_experimental_2004}
\APACinsertmetastar {%
baroud_experimental_2004}%
\begin{APACrefauthors}%
Baroud, G.%
, Falk, R.%
, Crookshank, M.%
, Sponagel, S.%
\BCBL {} Steffen, T.%
\end{APACrefauthors}%
\unskip\
\newblock
\APACrefYearMonthDay{2004}{}{}.
\newblock
{\BBOQ}\APACrefatitle {Experimental and theoretical investigation of
  directional permeability of human vertebral cancellous bone for cement
  infiltration} {Experimental and theoretical investigation of directional
  permeability of human vertebral cancellous bone for cement
  infiltration}.{\BBCQ}
\newblock
\APACjournalVolNumPages{J Biomech}{37}{2}{189--96}.
\newblock
\begin{APACrefURL} {https://doi.org/10.1016/S0021-9290(03)00246-X}
  \end{APACrefURL}
\newblock

\newblock

\PrintBackRefs{\CurrentBib}

\bibitem [\protect \citeauthoryear {%
Beaudoin%
, Mihalko%
\BCBL {}\ \BBA {} Krause%
}{%
Beaudoin%
\ \protect \BOthers {.}}{%
{\protect \APACyear {1991}}%
}]{%
beaudoin_finite_1991}
\APACinsertmetastar {%
beaudoin_finite_1991}%
\begin{APACrefauthors}%
Beaudoin, A.J.%
, Mihalko, W.M.%
\BCBL {} Krause, W.R.%
\end{APACrefauthors}%
\unskip\
\newblock
\APACrefYearMonthDay{1991}{}{}.
\newblock
{\BBOQ}\APACrefatitle {Finite element modelling of polymethylmethacrylate flow
  through cancellous bone} {Finite element modelling of polymethylmethacrylate
  flow through cancellous bone}.{\BBCQ}
\newblock
\APACjournalVolNumPages{J Biomech}{24}{2}{127--36}.
\newblock
\begin{APACrefURL} {https://doi.org/10.1016/0021-9290(91)90357-S}
  \end{APACrefURL}
\newblock

\newblock

\PrintBackRefs{\CurrentBib}

\bibitem [\protect \citeauthoryear {%
Bernhard%
}{%
Bernhard%
}{%
{\protect \APACyear {2003}}%
}]{%
bernhard_asymptomatic_2003}
\APACinsertmetastar {%
bernhard_asymptomatic_2003}%
\begin{APACrefauthors}%
Bernhard, J.%
\end{APACrefauthors}%
\unskip\
\newblock
\APACrefYearMonthDay{2003}{}{}.
\newblock
{\BBOQ}\APACrefatitle {Asymptomatic diffuse pulmonary embolism caused by
  acrylic cement: an unusual complication of percutaneous vertebroplasty}
  {Asymptomatic diffuse pulmonary embolism caused by acrylic cement: an unusual
  complication of percutaneous vertebroplasty}.{\BBCQ}
\newblock
\APACjournalVolNumPages{Ann Rheum Dis}{62}{1}{85--86}.
\newblock
\begin{APACrefURL} {https://doi.org/10.1136/ard.62.1.85} \end{APACrefURL}
\newblock

\newblock

\PrintBackRefs{\CurrentBib}

\bibitem [\protect \citeauthoryear {%
Bleiler%
\ \protect \BOthers {.}}{%
Bleiler%
\ \protect \BOthers {.}}{%
{\protect \APACyear {2015}}%
}]{%
bleiler_multiphasic_2015}
\APACinsertmetastar {%
bleiler_multiphasic_2015}%
\begin{APACrefauthors}%
Bleiler, C.%
, Wagner, A.%
, Stadelmann, V.A.%
, Windolf, M.%
, Köstler, H.%
, Boger, A.%
\BDBL {}Röhrle, O.%
\end{APACrefauthors}%
\unskip\
\newblock
\APACrefYearMonthDay{2015}{}{}.
\newblock
{\BBOQ}\APACrefatitle {Multiphasic modelling of bone-cement injection into
  vertebral cancellous bone} {Multiphasic modelling of bone-cement injection
  into vertebral cancellous bone}.{\BBCQ}
\newblock
\APACjournalVolNumPages{Int J Numer Method Biomed Eng}{31}{1}{423--43}.
\newblock
\begin{APACrefURL} {https://doi.org/10.1002/cnm.2696} \end{APACrefURL}
\newblock

\newblock

\PrintBackRefs{\CurrentBib}

\bibitem [\protect \citeauthoryear {%
Bohner%
, Gasser%
, Baroud%
\BCBL {}\ \BBA {} Heini%
}{%
Bohner%
\ \protect \BOthers {.}}{%
{\protect \APACyear {2003}}%
}]{%
bohner_theoretical_2003}
\APACinsertmetastar {%
bohner_theoretical_2003}%
\begin{APACrefauthors}%
Bohner, M.%
, Gasser, B.%
, Baroud, G.%
\BCBL {} Heini, P.%
\end{APACrefauthors}%
\unskip\
\newblock
\APACrefYearMonthDay{2003}{}{}.
\newblock
{\BBOQ}\APACrefatitle {Theoretical and experimental model to describe the
  injection of a polymethylmethacrylate cement into a porous structure}
  {Theoretical and experimental model to describe the injection of a
  polymethylmethacrylate cement into a porous structure}.{\BBCQ}
\newblock
\APACjournalVolNumPages{Biomaterials}{24}{16}{2721--30}.
\newblock
\begin{APACrefURL} {https://doi.org/10.1016/S0142-9612(03)00086-3}
  \end{APACrefURL}
\newblock

\newblock

\PrintBackRefs{\CurrentBib}

\bibitem [\protect \citeauthoryear {%
Bowen%
}{%
Bowen%
}{%
{\protect \APACyear {1984}}%
}]{%
bowen_porous_1984}
\APACinsertmetastar {%
bowen_porous_1984}%
\begin{APACrefauthors}%
Bowen, R.M.%
\end{APACrefauthors}%
\unskip\
\newblock
\APACrefYearMonthDay{1984}{}{}.
\newblock
{\BBOQ}\APACrefatitle {Porous {Media} {Model} {Formulations} by the {Theory} of
  {Mixtures}} {Porous {Media} {Model} {Formulations} by the {Theory} of
  {Mixtures}}.{\BBCQ}
\newblock
 J.~Bear\ \BBA {} M.Y.~Corapcioglu\ (\BEDS), \APACrefbtitle {Fundamentals of
  {Transport} {Phenomena} in {Porous} {Media}} {Fundamentals of {Transport}
  {Phenomena} in {Porous} {Media}}\ (\BPGS\ 63--119).
\newblock
\APACaddressPublisher{Dordrecht}{Springer Netherlands}.
\newblock
\begin{APACrefURL} {https://doi.org/10.1007/978-94-009-6175-3{\_}2}
  \end{APACrefURL}
\PrintBackRefs{\CurrentBib}

\bibitem [\protect \citeauthoryear {%
Brooks%
\ \BBA {} Corey%
}{%
Brooks%
\ \BBA {} Corey%
}{%
{\protect \APACyear {1964}}%
}]{%
brooks_corey_1964}
\APACinsertmetastar {%
brooks_corey_1964}%
\begin{APACrefauthors}%
Brooks, R.%
\BCBT {}\ \BBA {} Corey, A.%
\end{APACrefauthors}%
\unskip\
\newblock
\APACrefYear{1964}.
\newblock
\APACrefbtitle {Hydraulic Properties of Porous Media} {Hydraulic properties of
  porous media}.
\newblock
\APACaddressPublisher{Colorado State University, Fort Collins,
  Colorado}{Hydrology Papers}.
\PrintBackRefs{\CurrentBib}

\bibitem [\protect \citeauthoryear {%
Cannella%
\ \BBA {} Seright%
}{%
Cannella%
\ \BBA {} Seright%
}{%
{\protect \APACyear {1988}}%
}]{%
cannella_prediction_1988}
\APACinsertmetastar {%
cannella_prediction_1988}%
\begin{APACrefauthors}%
Cannella, W.J.%
\BCBT {}\ \BBA {} Seright, R.%
\end{APACrefauthors}%
\unskip\
\newblock
\APACrefYearMonthDay{1988}{}{}.
\newblock
{\BBOQ}\APACrefatitle {Prediction of {Xanthan} {Rheology} in {Porous} {Media}}
  {Prediction of {Xanthan} {Rheology} in {Porous} {Media}}.{\BBCQ}
\newblock
 \APACrefbtitle {SPE Annual Technical Conference and Exhibition} {Spe annual
  technical conference and exhibition}\ (\BVOL\ SPE 18089, \BPG~1-2).
\newblock
\APACaddressPublisher{Houston, TX}{Society of Petrolem Engineers}.
\PrintBackRefs{\CurrentBib}

\bibitem [\protect \citeauthoryear {%
Daish%
\ \protect \BOthers {.}}{%
Daish%
\ \protect \BOthers {.}}{%
{\protect \APACyear {2017}}%
}]{%
daish_estimation_2017}
\APACinsertmetastar {%
daish_estimation_2017}%
\begin{APACrefauthors}%
Daish, C.%
, Blanchard, R.%
, Gulati, K.%
, Losic, D.%
, Findlay, D.%
, Harvie, D.J.E.%
\BCBL {} Pivonka, P.%
\end{APACrefauthors}%
\unskip\
\newblock
\APACrefYearMonthDay{2017}{}{}.
\newblock
{\BBOQ}\APACrefatitle {Estimation of anisotropic permeability in trabecular
  bone based on {microCT} imaging and pore-scale fluid dynamics simulations}
  {Estimation of anisotropic permeability in trabecular bone based on {microCT}
  imaging and pore-scale fluid dynamics simulations}.{\BBCQ}
\newblock
\APACjournalVolNumPages{Bone Rep}{6}{}{129--39}.
\newblock
\begin{APACrefURL} {https://doi.org/10.1016/j.bonr.2016.12.002}
  \end{APACrefURL}
\newblock

\newblock

\PrintBackRefs{\CurrentBib}

\bibitem [\protect \citeauthoryear {%
Davis%
\ \BBA {} Praveen%
}{%
Davis%
\ \BBA {} Praveen%
}{%
{\protect \APACyear {2006}}%
}]{%
davis_nonlinear_2006}
\APACinsertmetastar {%
davis_nonlinear_2006}%
\begin{APACrefauthors}%
Davis, B.L.%
\BCBT {}\ \BBA {} Praveen, S.S.%
\end{APACrefauthors}%
\unskip\
\newblock
\APACrefYearMonthDay{2006}{}{}.
\newblock
{\BBOQ}\APACrefatitle {Nonlinear versus linear behavior of calcaneal bone
  marrow at different shear rates} {Nonlinear versus linear behavior of
  calcaneal bone marrow at different shear rates}.{\BBCQ}
\newblock
 \APACrefbtitle {Annual meeting of American Society of Biomechanics,
  Blacksburg, VA} {Annual meeting of american society of biomechanics,
  blacksburg, va}\ (\BPG~1-16).
\newblock
\APACaddressPublisher{Blacksburg, VA}{}.
\PrintBackRefs{\CurrentBib}

\bibitem [\protect \citeauthoryear {%
Eberhard%
\ \protect \BOthers {.}}{%
Eberhard%
\ \protect \BOthers {.}}{%
{\protect \APACyear {2019}}%
}]{%
eberhard_determination_2019}
\APACinsertmetastar {%
eberhard_determination_2019}%
\begin{APACrefauthors}%
Eberhard, U.%
, Seybold, H.J.%
, Floriancic, M.%
, Bertsch, P.%
, Jiménez-Martínez, J.%
, Andrade, J.S.J.%
\BCBL {} Holzner, M.%
\end{APACrefauthors}%
\unskip\
\newblock
\APACrefYearMonthDay{2019}{}{}.
\newblock
{\BBOQ}\APACrefatitle {Determination of the {Effective} {Viscosity} of
  {Non}-newtonian {Fluids} {Flowing} {Through} {Porous} {Media}} {Determination
  of the {Effective} {Viscosity} of {Non}-newtonian {Fluids} {Flowing}
  {Through} {Porous} {Media}}.{\BBCQ}
\newblock
\APACjournalVolNumPages{Front Phys}{7}{}{}.
\newblock
\begin{APACrefURL} {https://doi.org/10.3389/fphy.2019.00071} \end{APACrefURL}
\newblock

\newblock

\PrintBackRefs{\CurrentBib}

\bibitem [\protect \citeauthoryear {%
Ehlers%
}{%
Ehlers%
}{%
{\protect \APACyear {2002}}%
}]{%
ehlers_foundations_2002}
\APACinsertmetastar {%
ehlers_foundations_2002}%
\begin{APACrefauthors}%
Ehlers, W.%
\end{APACrefauthors}%
\unskip\
\newblock
\APACrefYearMonthDay{2002}{}{}.
\newblock
{\BBOQ}\APACrefatitle {Foundations of multiphasic and porous materials}
  {Foundations of multiphasic and porous materials}.{\BBCQ}
\newblock
\BIn{} W.~Ehlers\ \BBA {} J.~Bluhm\ (\BEDS), \APACrefbtitle {Porous Media:
  Theory, Experiments and Numerical Applications} {Porous media: Theory,
  experiments and numerical applications}\ (\BPGS\ 3--86).
\newblock
\APACaddressPublisher{Berlin, Heidelberg}{Springer Berlin Heidelberg}.
\newblock
\begin{APACrefURL} {https://doi.org/10.1007/978-3-662-04999-0{\_}1}
  \end{APACrefURL}
\PrintBackRefs{\CurrentBib}

\bibitem [\protect \citeauthoryear {%
Ehlers%
}{%
Ehlers%
}{%
{\protect \APACyear {2009}}%
}]{%
ehlers_challenges_2009}
\APACinsertmetastar {%
ehlers_challenges_2009}%
\begin{APACrefauthors}%
Ehlers, W.%
\end{APACrefauthors}%
\unskip\
\newblock
\APACrefYearMonthDay{2009}{}{}.
\newblock
{\BBOQ}\APACrefatitle {Challenges of porous media models in geo- and
  biomechanical engineering including electro-chemically active polymers and
  gels} {Challenges of porous media models in geo- and biomechanical
  engineering including electro-chemically active polymers and gels}.{\BBCQ}
\newblock
\APACjournalVolNumPages{Int J Adv Eng Sci Appl Math}{1}{1}{1--24}.
\newblock
\begin{APACrefURL} {https://doi.org/10.1007/s12572-009-0001-z} \end{APACrefURL}
\newblock

\newblock

\PrintBackRefs{\CurrentBib}

\bibitem [\protect \citeauthoryear {%
Ehlers%
, {Morrison Rehm}%
, Schr{\"o}der%
, St{\"o}hr%
\BCBL {}\ \BBA {} Wagner%
}{%
Ehlers%
\ \protect \BOthers {.}}{%
{\protect \APACyear {2022}}%
}]{%
Ehlers.2022}
\APACinsertmetastar {%
Ehlers.2022}%
\begin{APACrefauthors}%
Ehlers, W.%
, {Morrison Rehm}, M.%
, Schr{\"o}der, P.%
, St{\"o}hr, D.%
\BCBL {} Wagner, A.%
\end{APACrefauthors}%
\unskip\
\newblock
\APACrefYearMonthDay{2022}{}{}.
\newblock
{\BBOQ}\APACrefatitle {Multiphasic modelling and computation of metastatic
  lung-cancer cell proliferation and atrophy in brain tissue based on
  experimental data} {Multiphasic modelling and computation of metastatic
  lung-cancer cell proliferation and atrophy in brain tissue based on
  experimental data}.{\BBCQ}
\newblock
\APACjournalVolNumPages{Biomech Model Mechanobiol}{21}{1}{277--315}.
\newblock
\begin{APACrefURL} {https://doi.org/10.1007/s10237-021-01535-4}
  \end{APACrefURL}
\newblock

\newblock

\PrintBackRefs{\CurrentBib}

\bibitem [\protect \citeauthoryear {%
Gostick%
\ \protect \BOthers {.}}{%
Gostick%
\ \protect \BOthers {.}}{%
{\protect \APACyear {2019}}%
}]{%
gostick_porespy_2019}
\APACinsertmetastar {%
gostick_porespy_2019}%
\begin{APACrefauthors}%
Gostick, J.%
, Khan, Z.%
, Tranter, T.%
, Kok, M.%
, Agnaou, M.%
, Sadeghi, M.%
\BCBL {} Jervis, R.%
\end{APACrefauthors}%
\unskip\
\newblock
\APACrefYearMonthDay{2019}{}{}.
\newblock
{\BBOQ}\APACrefatitle {{PoreSpy}: {A} {Python} {Toolkit} for {Quantitative}
  {Analysis} of {Porous} {Media} {Images}} {{PoreSpy}: {A} {Python} {Toolkit}
  for {Quantitative} {Analysis} of {Porous} {Media} {Images}}.{\BBCQ}
\newblock
\APACjournalVolNumPages{J Open Source Softw}{4}{37}{1296}.
\newblock
\begin{APACrefURL} {https://doi.org/10.21105/joss.01296} \end{APACrefURL}
\newblock

\newblock

\PrintBackRefs{\CurrentBib}

\bibitem [\protect \citeauthoryear {%
Gurkan%
\ \BBA {} Akkus%
}{%
Gurkan%
\ \BBA {} Akkus%
}{%
{\protect \APACyear {2008}}%
}]{%
gurkan_mechanical_2008}
\APACinsertmetastar {%
gurkan_mechanical_2008}%
\begin{APACrefauthors}%
Gurkan, U.A.%
\BCBT {}\ \BBA {} Akkus, O.%
\end{APACrefauthors}%
\unskip\
\newblock
\APACrefYearMonthDay{2008}{}{}.
\newblock
{\BBOQ}\APACrefatitle {The {Mechanical} {Environment} of {Bone} {Marrow}: {A}
  {Review}} {The {Mechanical} {Environment} of {Bone} {Marrow}: {A}
  {Review}}.{\BBCQ}
\newblock
\APACjournalVolNumPages{Ann Biomed Eng}{36}{12}{1978--91}.
\newblock
\begin{APACrefURL} {https://doi.org/10.1007/s10439-008-9577-x} \end{APACrefURL}
\newblock

\newblock

\PrintBackRefs{\CurrentBib}

\bibitem [\protect \citeauthoryear {%
Hirasaki%
\ \BBA {} Pope%
}{%
Hirasaki%
\ \BBA {} Pope%
}{%
{\protect \APACyear {1974}}%
}]{%
hirasaki_analysis_1974}
\APACinsertmetastar {%
hirasaki_analysis_1974}%
\begin{APACrefauthors}%
Hirasaki, G.%
\BCBT {}\ \BBA {} Pope, G.%
\end{APACrefauthors}%
\unskip\
\newblock
\APACrefYearMonthDay{1974}{}{}.
\newblock
{\BBOQ}\APACrefatitle {Analysis of {Factors} {Influencing} {Mobility} and
  {Adsorption} in the {Flow} of {Polymer} {Solution} {Through} {Porous}
  {Media}} {Analysis of {Factors} {Influencing} {Mobility} and {Adsorption} in
  the {Flow} of {Polymer} {Solution} {Through} {Porous} {Media}}.{\BBCQ}
\newblock
\APACjournalVolNumPages{Soc Pet Eng J}{14}{04}{337--46}.
\newblock
\begin{APACrefURL} {https://doi.org/10.2118/4026-PA} \end{APACrefURL}
\newblock

\newblock

\PrintBackRefs{\CurrentBib}

\bibitem [\protect \citeauthoryear {%
Huber%
\ \BBA {} Helmig%
}{%
Huber%
\ \BBA {} Helmig%
}{%
{\protect \APACyear {2000}}%
}]{%
huber_node-centered_2000}
\APACinsertmetastar {%
huber_node-centered_2000}%
\begin{APACrefauthors}%
Huber, R.%
\BCBT {}\ \BBA {} Helmig, R.%
\end{APACrefauthors}%
\unskip\
\newblock
\APACrefYearMonthDay{2000}{}{}.
\newblock
{\BBOQ}\APACrefatitle {Node-centered ﬁnite volume discretizations for the
  numerical simulation of multiphase ﬂow in heterogeneous porous media}
  {Node-centered ﬁnite volume discretizations for the numerical simulation of
  multiphase ﬂow in heterogeneous porous media}.{\BBCQ}
\newblock
\APACjournalVolNumPages{Comput Geosci}{4}{}{141--64}.
\newblock
\begin{APACrefURL} {https://doi.org/10.1023/A:1011559916309} \end{APACrefURL}
\newblock

\newblock

\PrintBackRefs{\CurrentBib}

\bibitem [\protect \citeauthoryear {%
Jansen%
, Birch%
, Schiffman%
, Crosby%
\BCBL {}\ \BBA {} Peyton%
}{%
Jansen%
\ \protect \BOthers {.}}{%
{\protect \APACyear {2015}}%
}]{%
jansen_mechanics_2015}
\APACinsertmetastar {%
jansen_mechanics_2015}%
\begin{APACrefauthors}%
Jansen, L.E.%
, Birch, N.P.%
, Schiffman, J.D.%
, Crosby, A.J.%
\BCBL {} Peyton, S.R.%
\end{APACrefauthors}%
\unskip\
\newblock
\APACrefYearMonthDay{2015}{}{}.
\newblock
{\BBOQ}\APACrefatitle {Mechanics of {Intact} {Bone} {Marrow}} {Mechanics of
  {Intact} {Bone} {Marrow}}.{\BBCQ}
\newblock
\APACjournalVolNumPages{J Mech Behav Biomed Mater}{50}{}{299--307}.
\newblock
\begin{APACrefURL} {https://doi.org/10.1016/j.jmbbm.2015.06.023}
  \end{APACrefURL}
\newblock

\newblock

\PrintBackRefs{\CurrentBib}

\bibitem [\protect \citeauthoryear {%
Jensen%
\ \protect \BOthers {.}}{%
Jensen%
\ \protect \BOthers {.}}{%
{\protect \APACyear {1997}}%
}]{%
jensen_percutaneous_1997}
\APACinsertmetastar {%
jensen_percutaneous_1997}%
\begin{APACrefauthors}%
Jensen, M.E.%
, Evans, A.J.%
, Mathis, J.M.%
, Kallmes, D.F.%
, Cloft, H.J.%
\BCBL {} Dion, J.E.%
\end{APACrefauthors}%
\unskip\
\newblock
\APACrefYearMonthDay{1997}{}{}.
\newblock
{\BBOQ}\APACrefatitle {Percutaneous {Polymethylmethacrylate} {Vertebroplasty}
  in the {Treatment} of {Osteoporotic} {Vertebral} {Body} {Compression}
  {Fractures}: {Technical} {Aspects}} {Percutaneous {Polymethylmethacrylate}
  {Vertebroplasty} in the {Treatment} of {Osteoporotic} {Vertebral} {Body}
  {Compression} {Fractures}: {Technical} {Aspects}}.{\BBCQ}
\newblock
\APACjournalVolNumPages{Am J Neuroradiol}{18}{10}{1897--1904}.
\newblock

\newblock

\PrintBackRefs{\CurrentBib}

\bibitem [\protect \citeauthoryear {%
Joekar-Niasar%
, Hassanizadeh%
\BCBL {}\ \BBA {} Dahle%
}{%
Joekar-Niasar%
\ \protect \BOthers {.}}{%
{\protect \APACyear {2010}}%
}]{%
joekar-niasar_non-equilibrium_2010}
\APACinsertmetastar {%
joekar-niasar_non-equilibrium_2010}%
\begin{APACrefauthors}%
Joekar-Niasar, V.%
, Hassanizadeh, S.M.%
\BCBL {} Dahle, H.K.%
\end{APACrefauthors}%
\unskip\
\newblock
\APACrefYearMonthDay{2010}{}{}.
\newblock
{\BBOQ}\APACrefatitle {Non-equilibrium effects in capillarity and interfacial
  area in two-phase flow: dynamic pore-network modelling} {Non-equilibrium
  effects in capillarity and interfacial area in two-phase flow: dynamic
  pore-network modelling}.{\BBCQ}
\newblock
\APACjournalVolNumPages{J Fluid Mech}{655}{}{38--71}.
\newblock
\begin{APACrefURL} {https://doi.org/10.1017/S0022112010000704} \end{APACrefURL}
\newblock

\newblock

\PrintBackRefs{\CurrentBib}

\bibitem [\protect \citeauthoryear {%
Johnell%
\ \BBA {} Kanis%
}{%
Johnell%
\ \BBA {} Kanis%
}{%
{\protect \APACyear {2006}}%
}]{%
johnell_estimate_2006}
\APACinsertmetastar {%
johnell_estimate_2006}%
\begin{APACrefauthors}%
Johnell, O.%
\BCBT {}\ \BBA {} Kanis, J.A.%
\end{APACrefauthors}%
\unskip\
\newblock
\APACrefYearMonthDay{2006}{}{}.
\newblock
{\BBOQ}\APACrefatitle {An estimate of the worldwide prevalence and disability
  associated with osteoporotic fractures} {An estimate of the worldwide
  prevalence and disability associated with osteoporotic fractures}.{\BBCQ}
\newblock
\APACjournalVolNumPages{Osteoporos Int}{17}{12}{1726--33}.
\newblock
\begin{APACrefURL} {https://doi.org/10.1007/s00198-006-0172-4} \end{APACrefURL}
\newblock

\newblock

\PrintBackRefs{\CurrentBib}

\bibitem [\protect \citeauthoryear {%
Koch%
\ \protect \BOthers {.}}{%
Koch%
\ \protect \BOthers {.}}{%
{\protect \APACyear {2021}}%
}]{%
koch_dumux_2020}
\APACinsertmetastar {%
koch_dumux_2020}%
\begin{APACrefauthors}%
Koch, T.%
, Gläser, D.%
, Weishaupt, K.%
, Ackermann, S.%
, Beck, M.%
, Becker, B.%
\BDBL {}Flemisch, B.%
\end{APACrefauthors}%
\unskip\
\newblock
\APACrefYearMonthDay{2021}{}{}.
\newblock
{\BBOQ}\APACrefatitle {{DuMu}\textsuperscript{x} 3 - an open-source simulator
  for solving flow and transport problems in porous media with a focus on model
  coupling} {{DuMu}\textsuperscript{x} 3 - an open-source simulator for solving
  flow and transport problems in porous media with a focus on model
  coupling}.{\BBCQ}
\newblock
\APACjournalVolNumPages{Comput Math Appl}{81}{}{423-43}.
\newblock
\begin{APACrefURL} {https://doi.org/10.1016/j.camwa.2020.02.012}
  \end{APACrefURL}
\newblock

\newblock

\PrintBackRefs{\CurrentBib}

\bibitem [\protect \citeauthoryear {%
Krause%
, Miller%
\BCBL {}\ \BBA {} Ng%
}{%
Krause%
\ \protect \BOthers {.}}{%
{\protect \APACyear {1982}}%
}]{%
krause_viscosity_1982}
\APACinsertmetastar {%
krause_viscosity_1982}%
\begin{APACrefauthors}%
Krause, W.R.%
, Miller, J.%
\BCBL {} Ng, P.%
\end{APACrefauthors}%
\unskip\
\newblock
\APACrefYearMonthDay{1982}{}{}.
\newblock
{\BBOQ}\APACrefatitle {The viscosity of acrylic bone cements} {The viscosity of
  acrylic bone cements}.{\BBCQ}
\newblock
\APACjournalVolNumPages{J Biomed Mater Res}{16}{3}{219--243}.
\newblock
\begin{APACrefURL} [{2020-02-17}]{https://doi.org/10.1002/jbm.820160305}
  \end{APACrefURL}
\newblock

\newblock

\PrintBackRefs{\CurrentBib}

\bibitem [\protect \citeauthoryear {%
Landgraf%
\ \protect \BOthers {.}}{%
Landgraf%
\ \protect \BOthers {.}}{%
{\protect \APACyear {2015}}%
}]{%
landgraf_modelling_2015}
\APACinsertmetastar {%
landgraf_modelling_2015}%
\begin{APACrefauthors}%
Landgraf, R.%
, Ihlemann, J.%
, Kolmeder, S.%
, Lion, A.%
, Lebsack, H.%
\BCBL {} Kober, C.%
\end{APACrefauthors}%
\unskip\
\newblock
\APACrefYearMonthDay{2015}{}{}.
\newblock
{\BBOQ}\APACrefatitle {Modelling and simulation of acrylic bone cement
  injection and curing within the framework of vertebroplasty} {Modelling and
  simulation of acrylic bone cement injection and curing within the framework
  of vertebroplasty}.{\BBCQ}
\newblock
\APACjournalVolNumPages{Z Angew Math Mech}{95}{12}{1530--47}.
\newblock
\begin{APACrefURL} {https://doi.org/10.1002/zamm.201400064} \end{APACrefURL}
\newblock

\newblock

\PrintBackRefs{\CurrentBib}

\bibitem [\protect \citeauthoryear {%
Lepoutre%
, Meylheuc%
, Bara%
, Barbé%
\BCBL {}\ \BBA {} Bayle%
}{%
Lepoutre%
\ \protect \BOthers {.}}{%
{\protect \APACyear {2019}}%
}]{%
lepoutre_bone_2019}
\APACinsertmetastar {%
lepoutre_bone_2019}%
\begin{APACrefauthors}%
Lepoutre, N.%
, Meylheuc, L.%
, Bara, G.I.%
, Barbé, L.%
\BCBL {} Bayle, B.%
\end{APACrefauthors}%
\unskip\
\newblock
\APACrefYearMonthDay{2019}{}{}.
\newblock
{\BBOQ}\APACrefatitle {Bone cement modeling for percutaneous vertebroplasty}
  {Bone cement modeling for percutaneous vertebroplasty}.{\BBCQ}
\newblock
\APACjournalVolNumPages{J Biomed Mater Res - B Appl
  Biomater}{107}{5}{1504--1515}.
\newblock
\begin{APACrefURL} [{2020-03-12}]{https://doi.org/10.1002/jbm.b.34242}
  \end{APACrefURL}
\newblock

\newblock

\PrintBackRefs{\CurrentBib}

\bibitem [\protect \citeauthoryear {%
Lian%
, Chui%
\BCBL {}\ \BBA {} Teoh%
}{%
Lian%
\ \protect \BOthers {.}}{%
{\protect \APACyear {2008}}%
}]{%
lian_biomechanical_2008}
\APACinsertmetastar {%
lian_biomechanical_2008}%
\begin{APACrefauthors}%
Lian, Z.%
, Chui, C\BHBI K.%
\BCBL {} Teoh, S\BHBI H.%
\end{APACrefauthors}%
\unskip\
\newblock
\APACrefYearMonthDay{2008}{}{}.
\newblock
{\BBOQ}\APACrefatitle {A biomechanical model for real-time simulation of {PMMA}
  injection with haptics} {A biomechanical model for real-time simulation of
  {PMMA} injection with haptics}.{\BBCQ}
\newblock
\APACjournalVolNumPages{Comput Biol Med}{38}{3}{304--312}.
\newblock
\begin{APACrefURL} {https://doi.org/10.1016/j.compbiomed.2007.10.009}
  \end{APACrefURL}
\newblock

\newblock

\PrintBackRefs{\CurrentBib}

\bibitem [\protect \citeauthoryear {%
Loeffel%
, Ferguson%
, Nolte%
\BCBL {}\ \BBA {} Kowal%
}{%
Loeffel%
\ \protect \BOthers {.}}{%
{\protect \APACyear {2008}}%
}]{%
loeffel_vertebroplasty_2008}
\APACinsertmetastar {%
loeffel_vertebroplasty_2008}%
\begin{APACrefauthors}%
Loeffel, M.%
, Ferguson, S.J.%
, Nolte, L.P.%
\BCBL {} Kowal, J.H.%
\end{APACrefauthors}%
\unskip\
\newblock
\APACrefYearMonthDay{2008}{}{}.
\newblock
{\BBOQ}\APACrefatitle {Vertebroplasty: {Experimental} {Characterization} of
  {Polymethylmethacrylate} {Bone} {Cement} {Spreading} as a {Function} of
  {Viscosity}, {Bone} {Porosity}, and {Flow} {Rate}} {Vertebroplasty:
  {Experimental} {Characterization} of {Polymethylmethacrylate} {Bone} {Cement}
  {Spreading} as a {Function} of {Viscosity}, {Bone} {Porosity}, and {Flow}
  {Rate}}.{\BBCQ}
\newblock
\APACjournalVolNumPages{Spine}{33}{12}{1352--1359}.
\newblock
\begin{APACrefURL} {https://doi.org/10.1097/BRS.0b013e3181732aa9}
  \end{APACrefURL}
\newblock

\newblock

\PrintBackRefs{\CurrentBib}

\bibitem [\protect \citeauthoryear {%
Markert%
}{%
Markert%
}{%
{\protect \APACyear {2007}}%
}]{%
markert_constitutive_2007}
\APACinsertmetastar {%
markert_constitutive_2007}%
\begin{APACrefauthors}%
Markert, B.%
\end{APACrefauthors}%
\unskip\
\newblock
\APACrefYearMonthDay{2007}{}{}.
\newblock
{\BBOQ}\APACrefatitle {A constitutive approach to 3-d nonlinear fluid flow
  through finite deformable porous continua: {With} application to a
  high-porosity polyurethane foam} {A constitutive approach to 3-d nonlinear
  fluid flow through finite deformable porous continua: {With} application to a
  high-porosity polyurethane foam}.{\BBCQ}
\newblock
\APACjournalVolNumPages{Transp Porous Med}{70}{3}{427--450}.
\newblock
\begin{APACrefURL} {https://doi.org/10.1007/s11242-007-9107-6} \end{APACrefURL}
\newblock

\newblock

\PrintBackRefs{\CurrentBib}

\bibitem [\protect \citeauthoryear {%
McGraw%
\ \protect \BOthers {.}}{%
McGraw%
\ \protect \BOthers {.}}{%
{\protect \APACyear {2002}}%
}]{%
mcgraw_prospective_2002}
\APACinsertmetastar {%
mcgraw_prospective_2002}%
\begin{APACrefauthors}%
McGraw, J.K.%
, Lippert, J.A.%
, Minkus, K.D.%
, Rami, P.M.%
, Davis, T.M.%
\BCBL {} Budzik, R.F.%
\end{APACrefauthors}%
\unskip\
\newblock
\APACrefYearMonthDay{2002}{}{}.
\newblock
{\BBOQ}\APACrefatitle {Prospective evaluation of pain relief in 100 patients
  undergoing percutaneous vertebroplasty: results and follow-up} {Prospective
  evaluation of pain relief in 100 patients undergoing percutaneous
  vertebroplasty: results and follow-up}.{\BBCQ}
\newblock
\APACjournalVolNumPages{J Vasc Interv Radiol}{13}{9 Pt 1}{883--86}.
\newblock
\begin{APACrefURL} {https://doi.org/10.1016/s1051-0443(07)61770-9}
  \end{APACrefURL}
\newblock

\newblock

\PrintBackRefs{\CurrentBib}

\bibitem [\protect \citeauthoryear {%
Metzger%
, Shudick%
, Seekell%
, Zhu%
\BCBL {}\ \BBA {} Niebur%
}{%
Metzger%
\ \protect \BOthers {.}}{%
{\protect \APACyear {2014}}%
}]{%
metzger_rheological_2014}
\APACinsertmetastar {%
metzger_rheological_2014}%
\begin{APACrefauthors}%
Metzger, T.A.%
, Shudick, J.M.%
, Seekell, R.%
, Zhu, Y.%
\BCBL {} Niebur, G.L.%
\end{APACrefauthors}%
\unskip\
\newblock
\APACrefYearMonthDay{2014}{}{}.
\newblock
{\BBOQ}\APACrefatitle {Rheological behavior of fresh bone marrow and the
  effects of storage} {Rheological behavior of fresh bone marrow and the
  effects of storage}.{\BBCQ}
\newblock
\APACjournalVolNumPages{J Mech Behav Biomed Mater}{40}{}{307--13}.
\newblock
\begin{APACrefURL} {https://doi.org/10.1016/j.jmbbm.2014.09.008}
  \end{APACrefURL}
\newblock

\newblock

\PrintBackRefs{\CurrentBib}

\bibitem [\protect \citeauthoryear {%
Ochia%
\ \BBA {} Ching%
}{%
Ochia%
\ \BBA {} Ching%
}{%
{\protect \APACyear {2002}}%
}]{%
ochia_hydraulic_2002}
\APACinsertmetastar {%
ochia_hydraulic_2002}%
\begin{APACrefauthors}%
Ochia, R.S.%
\BCBT {}\ \BBA {} Ching, R.P.%
\end{APACrefauthors}%
\unskip\
\newblock
\APACrefYearMonthDay{2002}{}{}.
\newblock
{\BBOQ}\APACrefatitle {Hydraulic {Resistance} and {Permeability} in {Human}
  {Lumbar} {Vertebral} {Bodies}} {Hydraulic {Resistance} and {Permeability} in
  {Human} {Lumbar} {Vertebral} {Bodies}}.{\BBCQ}
\newblock
\APACjournalVolNumPages{J Biomech Eng}{124}{5}{533--37}.
\newblock
\begin{APACrefURL} {https://doi.org/10.1115/1.1503793} \end{APACrefURL}
\newblock

\newblock

\PrintBackRefs{\CurrentBib}

\bibitem [\protect \citeauthoryear {%
Ratliff%
, Nguyen%
\BCBL {}\ \BBA {} Heiss%
}{%
Ratliff%
\ \protect \BOthers {.}}{%
{\protect \APACyear {2001}}%
}]{%
ratliff_root_2001}
\APACinsertmetastar {%
ratliff_root_2001}%
\begin{APACrefauthors}%
Ratliff, J.%
, Nguyen, T.%
\BCBL {} Heiss, J.%
\end{APACrefauthors}%
\unskip\
\newblock
\APACrefYearMonthDay{2001}{}{}.
\newblock
{\BBOQ}\APACrefatitle {Root and {Spinal} {Cord} {Compression} from
  {Methylmethacrylate} {Vertebroplasty}:} {Root and {Spinal} {Cord}
  {Compression} from {Methylmethacrylate} {Vertebroplasty}:}.{\BBCQ}
\newblock
\APACjournalVolNumPages{Spine J}{26}{13}{e300--e302}.
\newblock
\begin{APACrefURL} {https://doi.org/10.1097/00007632-200107010-00021}
  \end{APACrefURL}
\newblock

\newblock

\PrintBackRefs{\CurrentBib}

\bibitem [\protect \citeauthoryear {%
Ricken%
\ \BBA {} Lambers%
}{%
Ricken%
\ \BBA {} Lambers%
}{%
{\protect \APACyear {2019}}%
}]{%
Ricken.2019}
\APACinsertmetastar {%
Ricken.2019}%
\begin{APACrefauthors}%
Ricken, T.%
\BCBT {}\ \BBA {} Lambers, L.%
\end{APACrefauthors}%
\unskip\
\newblock
\APACrefYearMonthDay{2019}{}{}.
\newblock
{\BBOQ}\APACrefatitle {On computational approaches of liver lobule function and
  perfusion simulation} {On computational approaches of liver lobule function
  and perfusion simulation}.{\BBCQ}
\newblock
\APACjournalVolNumPages{GAMM-Mitteilungen}{42}{4}{e201900016}.
\newblock
\begin{APACrefURL} {https://doi.org/10.1002/gamm.201900016} \end{APACrefURL}
\newblock

\newblock

\PrintBackRefs{\CurrentBib}

\bibitem [\protect \citeauthoryear {%
Sochi%
}{%
Sochi%
}{%
{\protect \APACyear {2015}}%
}]{%
sochi_analytical_2015}
\APACinsertmetastar {%
sochi_analytical_2015}%
\begin{APACrefauthors}%
Sochi, T.%
\end{APACrefauthors}%
\unskip\
\newblock
\APACrefYearMonthDay{2015}{}{}.
\newblock
{\BBOQ}\APACrefatitle {Analytical solutions for the flow of {Carreau} and
  {Cross} fluids in circular pipes and thin slits} {Analytical solutions for
  the flow of {Carreau} and {Cross} fluids in circular pipes and thin
  slits}.{\BBCQ}
\newblock
\APACjournalVolNumPages{Rheol Acta}{54}{8}{745--756}.
\newblock
\begin{APACrefURL} [{2022-05-13}]{https://doi.org/10.1007/s00397-015-0863-x}
  \end{APACrefURL}
\newblock

\newblock

\PrintBackRefs{\CurrentBib}

\bibitem [\protect \citeauthoryear {%
J.~Teo%
, Wang%
\BCBL {}\ \BBA {} Teoh%
}{%
J.~Teo%
\ \protect \BOthers {.}}{%
{\protect \APACyear {2007}}%
}]{%
teo_preliminary_2007}
\APACinsertmetastar {%
teo_preliminary_2007}%
\begin{APACrefauthors}%
Teo, J.%
, Wang, S.C.%
\BCBL {} Teoh, S.H.%
\end{APACrefauthors}%
\unskip\
\newblock
\APACrefYearMonthDay{2007}{}{}.
\newblock
{\BBOQ}\APACrefatitle {Preliminary {Study} on {Biomechanics} of
  {Vertebroplasty}: {A} {Computational} {Fluid} {Dynamics} and {Solid}
  {Mechanics} {Combined} {Approach}} {Preliminary {Study} on {Biomechanics} of
  {Vertebroplasty}: {A} {Computational} {Fluid} {Dynamics} and {Solid}
  {Mechanics} {Combined} {Approach}}.{\BBCQ}
\newblock
\APACjournalVolNumPages{Spine J}{32}{12}{1320--28}.
\newblock
\begin{APACrefURL} {https://doi.org/10.1097/BRS.0b013e318059af56}
  \end{APACrefURL}
\newblock

\newblock

\PrintBackRefs{\CurrentBib}

\bibitem [\protect \citeauthoryear {%
J.C.M.~Teo%
, Si-Hoe%
, Keh%
\BCBL {}\ \BBA {} Teoh%
}{%
J.C.M.~Teo%
\ \protect \BOthers {.}}{%
{\protect \APACyear {2007}}%
}]{%
teo_correlation_2007}
\APACinsertmetastar {%
teo_correlation_2007}%
\begin{APACrefauthors}%
Teo, J.C.M.%
, Si-Hoe, K.M.%
, Keh, J.E.L.%
\BCBL {} Teoh, S.H.%
\end{APACrefauthors}%
\unskip\
\newblock
\APACrefYearMonthDay{2007}{}{}.
\newblock
{\BBOQ}\APACrefatitle {Correlation of cancellous bone microarchitectural
  parameters from {microCT} to {CT} number and bone mechanical properties}
  {Correlation of cancellous bone microarchitectural parameters from {microCT}
  to {CT} number and bone mechanical properties}.{\BBCQ}
\newblock
\APACjournalVolNumPages{Mater Sci Eng C}{27}{2}{333--339}.
\newblock
\begin{APACrefURL} {https://doi.org/10.1016/j.msec.2006.05.003}
  \end{APACrefURL}
\newblock

\newblock

\PrintBackRefs{\CurrentBib}

\bibitem [\protect \citeauthoryear {%
Trivedi%
\ \protect \BOthers {.}}{%
Trivedi%
\ \protect \BOthers {.}}{%
{\protect \APACyear {2022}}%
}]{%
darus-3146_2022}
\APACinsertmetastar {%
darus-3146_2022}%
\begin{APACrefauthors}%
Trivedi, Z.%
, Gehweiler, D.%
, Wychowaniec, J.%
, Ricken, T.%
, Gueorguiev-Rüegg, B.%
, Wagner, A.%
\BCBL {} Röhrle, O.%
\end{APACrefauthors}%
\unskip\
\newblock
\APACrefYearMonthDay{2022}{}{}.
\newblock
\APACrefbtitle {{Data for: A continuum mechanical porous media model for
  simulating vertebroplasty: Numerical simulations and experimental
  validation}.} {{Data for: A continuum mechanical porous media model for
  simulating vertebroplasty: Numerical simulations and experimental
  validation}.}
\newblock
\APACaddressPublisher{}{DaRUS}.
\newblock
\begin{APACrefURL} {https://doi.org/10.18419/darus-3146} \end{APACrefURL}
\PrintBackRefs{\CurrentBib}

\bibitem [\protect \citeauthoryear {%
Wang%
, Eriksson%
, Ricken%
\BCBL {}\ \BBA {} Pierce%
}{%
Wang%
\ \protect \BOthers {.}}{%
{\protect \APACyear {2018}}%
}]{%
Wang.2018}
\APACinsertmetastar {%
Wang.2018}%
\begin{APACrefauthors}%
Wang, X.%
, Eriksson, T.S.E.%
, Ricken, T.%
\BCBL {} Pierce, D.M.%
\end{APACrefauthors}%
\unskip\
\newblock
\APACrefYearMonthDay{2018}{}{}.
\newblock
{\BBOQ}\APACrefatitle {On incorporating osmotic prestretch/prestress in
  image-driven finite element simulations of cartilage} {On incorporating
  osmotic prestretch/prestress in image-driven finite element simulations of
  cartilage}.{\BBCQ}
\newblock
\APACjournalVolNumPages{J Mech Behav Biomed Mater}{86}{}{409--422}.
\newblock
\begin{APACrefURL} {https://doi.org/10.1016/j.jmbbm.2018.06.014}
  \end{APACrefURL}
\newblock

\newblock

\PrintBackRefs{\CurrentBib}

\bibitem [\protect \citeauthoryear {%
Widmer%
}{%
Widmer%
}{%
{\protect \APACyear {2011}}%
}]{%
widmer_mixed_2011}
\APACinsertmetastar {%
widmer_mixed_2011}%
\begin{APACrefauthors}%
Widmer, R.P.%
\end{APACrefauthors}%
\unskip\
\newblock
\APACrefYearMonthDay{2011}{}{}.
\newblock
{\BBOQ}\APACrefatitle {A {Mixed} {Boundary} {Representation} to {Simulate} the
  {Displacement} of a {Biofluid} by a {Biomaterial} in {Porous} {Media}} {A
  {Mixed} {Boundary} {Representation} to {Simulate} the {Displacement} of a
  {Biofluid} by a {Biomaterial} in {Porous} {Media}}.{\BBCQ}
\newblock
\APACjournalVolNumPages{J Biomech Eng}{133}{5}{051007}.
\newblock
\begin{APACrefURL} {https://doi.org/10.1115/1.4003735} \end{APACrefURL}
\newblock

\newblock

\PrintBackRefs{\CurrentBib}

\bibitem [\protect \citeauthoryear {%
Widmer~Soyka%
, López%
, Persson%
, Cristofolini%
\BCBL {}\ \BBA {} Ferguson%
}{%
Widmer~Soyka%
\ \protect \BOthers {.}}{%
{\protect \APACyear {2013}}%
}]{%
widmer_soyka_numerical_2013}
\APACinsertmetastar {%
widmer_soyka_numerical_2013}%
\begin{APACrefauthors}%
Widmer~Soyka, R.P.%
, López, A.%
, Persson, C.%
, Cristofolini, L.%
\BCBL {} Ferguson, S.J.%
\end{APACrefauthors}%
\unskip\
\newblock
\APACrefYearMonthDay{2013}{}{}.
\newblock
{\BBOQ}\APACrefatitle {Numerical description and experimental validation of a
  rheology model for non-{Newtonian} fluid flow in cancellous bone} {Numerical
  description and experimental validation of a rheology model for
  non-{Newtonian} fluid flow in cancellous bone}.{\BBCQ}
\newblock
\APACjournalVolNumPages{J Mech Behav Biomed Mater}{27}{}{43--53}.
\newblock
\begin{APACrefURL} {https://doi.org/10.1016/j.jmbbm.2013.06.007}
  \end{APACrefURL}
\newblock

\newblock

\PrintBackRefs{\CurrentBib}

\bibitem [\protect \citeauthoryear {%
Wu%
, Isaksson%
, Ferguson%
\BCBL {}\ \BBA {} Persson%
}{%
Wu%
\ \protect \BOthers {.}}{%
{\protect \APACyear {2018}}%
}]{%
wu_youngs_2018}
\APACinsertmetastar {%
wu_youngs_2018}%
\begin{APACrefauthors}%
Wu, D.%
, Isaksson, P.%
, Ferguson, S.J.%
\BCBL {} Persson, C.%
\end{APACrefauthors}%
\unskip\
\newblock
\APACrefYearMonthDay{2018}{}{}.
\newblock
{\BBOQ}\APACrefatitle {Young’s modulus of trabecular bone at the tissue
  level: {A} review} {Young’s modulus of trabecular bone at the tissue level:
  {A} review}.{\BBCQ}
\newblock
\APACjournalVolNumPages{Acta Biomater}{78}{}{1--12}.
\newblock
\begin{APACrefURL} {https://doi.org/10.1016/j.actbio.2018.08.001}
  \end{APACrefURL}
\newblock

\newblock

\PrintBackRefs{\CurrentBib}

\bibitem [\protect \citeauthoryear {%
Zeiser%
, Bashoor-Zadeh%
, Darabi%
\BCBL {}\ \BBA {} Baroud%
}{%
Zeiser%
\ \protect \BOthers {.}}{%
{\protect \APACyear {2008}}%
}]{%
zeiser_pore-scale_2008}
\APACinsertmetastar {%
zeiser_pore-scale_2008}%
\begin{APACrefauthors}%
Zeiser, T.%
, Bashoor-Zadeh, M.%
, Darabi, A.%
\BCBL {} Baroud, G.%
\end{APACrefauthors}%
\unskip\
\newblock
\APACrefYearMonthDay{2008}{}{}.
\newblock
{\BBOQ}\APACrefatitle {Pore-scale analysis of {Newtonian} flow in the explicit
  geometry of vertebral trabecular bones using lattice {Boltzmann} simulation}
  {Pore-scale analysis of {Newtonian} flow in the explicit geometry of
  vertebral trabecular bones using lattice {Boltzmann} simulation}.{\BBCQ}
\newblock
\APACjournalVolNumPages{Proc Inst Mech Eng, Part H: J Eng
  Med}{222}{2}{185--94}.
\newblock
\begin{APACrefURL} {https://doi.org/10.1243/09544119JEIM261} \end{APACrefURL}
\newblock

\newblock

\PrintBackRefs{\CurrentBib}

\bibitem [\protect \citeauthoryear {%
Zienkiewicz%
\ \BBA {} Heinrich%
}{%
Zienkiewicz%
\ \BBA {} Heinrich%
}{%
{\protect \APACyear {1978}}%
}]{%
zienkiewicz_finite_1978}
\APACinsertmetastar {%
zienkiewicz_finite_1978}%
\begin{APACrefauthors}%
Zienkiewicz, O.%
\BCBT {}\ \BBA {} Heinrich, J.%
\end{APACrefauthors}%
\unskip\
\newblock
\APACrefYearMonthDay{1978}{}{}.
\newblock
{\BBOQ}\APACrefatitle {The finite element method and convection problems in
  fluid mechanics} {The finite element method and convection problems in fluid
  mechanics}.{\BBCQ}
\newblock
 R.~Gallagher, O.~Zienkiewicz, J.~Oden, M.~Morandi~Cecchi\BCBL {}\ \BBA {}
  C.~Taylor\ (\BEDS), \APACrefbtitle {Finite Element in Fluids} {Finite element
  in fluids}\ (\BVOL~3).
\newblock
\APACaddressPublisher{London}{Wiley}.
\PrintBackRefs{\CurrentBib}

\end{thebibliography}


\end{document}